\documentclass{aa501}
\usepackage{times}
\usepackage{latexsym}
\usepackage{epsfig}

\def\pho{\ifmmode\phantom{0}\else$\phantom{0}$\fi}
\def\kms{\ifmmode{\rm km\,s^{-1}}\else$\rm km\,s^{-1}$\fi}

\def\VelaX1{\hbox{Vela X-1}}
\makeatletter
\def\ion#1#2{#1$\;${\sc\@roman{#2}}\relax}
\makeatother
\begin{document}

\title{The mass of the neutron star in Vela X-1\thanks{Based 
on observations collected at the European Southern Observatory, Chile
(56.D-0251, 57.D-0409).}}

\author{O.~Barziv\inst{1,2} 
   \and L.~Kaper\inst{2}
   \and M.H.~van~Kerkwijk\inst{3}
   \and J.H.~Telting\inst{4} 
   \and J.~van~Paradijs\inst{2,5}\thanks{Deceased 1999 November 2}}

\offprints{L. Kaper (lexk@astro.uva.nl)}

\institute{European Southern Observatory, Karl-Schwarzschild-Strasse 2,
85748 Garching, Germany 
\and Astronomical Institute ``Anton Pannekoek'', University of
Amsterdam and Center for High-Energy Astrophysics, \\
Kruislaan 403, 1098 SJ Amsterdam, The Netherlands
\and Astronomical Institute, Utrecht University, P. O. Box 80000,
3508 TA Utrecht, The Netherlands  
\and Isaac Newton Group of Telescopes, Netherlands Organisation for
Scientific Research (NWO), Apartado 321, E-38700 Santa
Cruz de La Palma, Spain
\and Physics Department, University of Alabama, Huntsville, AL35899,
USA}

\date{Received; Accepted}

\markboth{O.\ Barziv et al.: The mass of Vela X-1}{}

\abstract{We measured the radial-velocity curve of HD~77581, the
B-supergiant companion of the X-ray pulsar \VelaX1, using 183
high-resolution optical spectra obtained in a nine-month campaign.  We
derive radial-velocity amplitudes for different lines and wavelength
regions, and find all are consistent with each other, as well as with
values found in previous analyses.  We show that one apparent
exception, an anomalously low value derived from ultra-violet spectra
obtained with the {\it International Ultraviolet Explorer}, was due
to an error in the analysis procedures.  We re-analyse all IUE
spectra, and combine the resulting velocities with the ones derived
from the new optical spectra presented here, as well as those derived
from optical spectra published earlier.  As in previous analyses, the
radial velocities show strong deviations from those expected for a
pure Keplerian orbit, with root-mean-square amplitudes of
$\sim\!7~\kms$ for strong lines of \ion{Si}{4} and \ion{N}{3} near
4100~\AA, and up to $\sim\!20~\kms$ for weaker lines of \ion{N}{2} and
\ion{Al}{3} near 5700~\AA.  The deviations likely are related to the
pronounced line-profile variations seen in our spectra.  Our hope was
that the deviations would average out when a sufficient number of
spectra were added together.  It turns out, however, that systematic
deviations as a function of orbital phase are present as well, at the
3~\kms\ level, with the largest deviations occurring near inferior
conjunction of the neutron star and near the phase of maximum
approaching velocity.  While the former might be due to a
photo-ionisation wake, for which we observe direct evidence in the
profiles of H$\delta$ and H$\alpha$, the latter has no straightforward
explanation.  As a result, our best estimate of the radial-velocity
amplitude, $K_{\rm opt}=21.7\pm1.6$~\kms, has an uncertainty not much
reduced to that found in previous analyses, in which the influence of
the systematic, phase-locked deviations had not been taken into
account.  Combining our velocity amplitude with the accurate orbital
elements of the X-ray pulsar, we infer $M_{\rm
ns}\sin^3{i}=1.78\pm0.15$~$M_\odot$\thanks{The tables in the Appendix
are only available in electronic form at the CDS via anonymous ftp to
cdsarc.u-strasbg.fr (130.79.128.5) or via
http://cdsweb.u-strasbg.fr/cgi-bin/qcat?/A+A/}.
}

\maketitle

\section{Introduction}\label{sec:intro}

Neutron stars follow a mass-radius relation that depends on the
equation of state of the superdense matter in their interiors.  For
densities up to the nuclear saturation density
($\sim\!10^{14.5}{\rm~g\,cm^{-3}}$), the equation of state is
reasonably well constrained by laboratory measurements, but for higher
densities it is still under much debate.  For a given equation of
state, there is a maximum possible mass of a neutron star.  For a
``soft'' one (i.e., ``compressible'' matter), the maximum is near
1.5~$M_{\odot}$, while for a ``stiff'' one (i.e., incompressible
matter) it can be as high as 2~$M_{\odot}$ (for recent reviews, see
Heiselberg \& Pandharipande \cite{HP00}; Lattimer \& Prakash
\cite{LP00}; Srinivasan \cite{Sr01}).  In this respect, the X-ray
pulsar \VelaX1 is of particular interest, as it is the only neutron
star for which there is strong evidence that it is more massive than
the canonical 1.4~$M_{\odot}$ (for a review, see Van Kerkwijk et al.\
\cite{vKvPZ95}; Thorsett \& Chakrabarthy \cite{TC99}).

Vela X-1 (4U\,0900--40) is a 283-s pulsar orbiting the B0.5~Ib
supergiant HD~77581 in a period of 8.96 days.  The X-ray source is
powered by accretion from the B supergiant's wind (see Kaper
\cite{Ka01} for a recent overview of high-mass X-ray binaries hosting
an OB-supergiant). Since the discovery of the X-ray pulsations in
\VelaX1 by McClintock et al.\ (\cite{McCRJ+76}), its slightly
eccentric ($e=0.09$) orbit has been determined with increasing
accuracy from Doppler-shift measurements of pulse-arrival times.
Recently, an update of the orbital parameters has been obtained based
on the five-year long series of frequency measurements in hard X-ray
flux with the Burst And Transient Source Experiment (BATSE) on board
the {\it Compton Gamma-Ray Observatory} (Bildsten et al.\
\cite{BCC+97}); see Table~\ref{tab:batse}.

The evidence for \VelaX1's high mass is the relatively large
amplitude, $K_{\rm{}opt}$, of the radial-velocity variations of the
B~supergiant companion, first noticed by Van Paradijs et al.\
(\cite{vPH76}).  From their analysis of close to a hundred
photographic spectra, Van Paradijs et al.\ (\cite{vPZT+77}, hereafter
Paper~I) found $K_{\rm{}opt}=21.75\pm1.15~\kms$, which, combined with
the BATSE orbit, corresponds to a neutron-star mass of
$M_{\rm{}ns}\sin^3{i}=1.79\pm0.11~M_{\odot}$ (here $M_{\rm{}ns}$ is
the mass of the neutron star, and $i$ is the orbital inclination;
uncertainties are 1$\sigma$ unless specified otherwise).  Guided by
the prospect that a more accurate mass determination of \VelaX1 might
provide an important constraint on the equation of state appropriate
for matter in neutron stars, Van Kerkwijk et al.\ (\cite{vKvPZ+95},
Paper~II) made a detailed study of the radial-velocity curve of
HD~77581, based on 40 high signal-to-noise CCD spectra, as well as 26
ultraviolet spectra obtained with the {\em International Ultraviolet
Explorer} (IUE) and 13 nightly averages of digitized
photographic spectra. They derived $K_{\rm{}opt}=20.8\pm1.4~\kms$,
confirming the value found in Paper~I.

In Paper~II, the accuracy of $K_{\rm{}opt}$ was found to be limited
not by the statistical uncertainty in the radial-velocity
measurements, but by systematic non-orbital contributions to the
radial velocity.  These contributions are correlated within a given
night and can reach amplitudes up to $\sim\!15~\kms$.  They did not
appear to be correlated with orbital phase.  It was suggested that
these velocity deviations might be caused by non-radial pulsations of
HD~77581, similar to those seen in many early-type stars, perhaps
excited by the varying tidal force excerted by the neutron star in its
eccentric orbit.  The situation may be complicated further by
systematic effects on the spectra due to a photo-ionisation wake
trailing the neutron star in its orbit (Kaper et al.\ \cite{KHZ94}).

In Paper~II, it was suggested that for a better understanding of these
radial-velocity excursions, and thereby an improvement of the accuracy
of the radial-velocity curve, either an extensive long-term sequence
of spectroscopic observations would be required, with which one might
``average out'' the deviations, or a continuous spectroscopy campaign
with a time resolution of a few hours, extending over at least one
orbital cycle.  We followed the first approach and monitored the
binary system for a period of about 9 months, taking high-resolution
spectra of HD~77581 on a night-by-night basis at the European Southern
Observatory.  

In Sect.~\ref{sec:reduction}, we describe the observations and the
data-reduction procedures. In Sect.~\ref{sec:spectra}, the spectra
are presented. In Sects.~\ref{sec:radvel} and~\ref{sec:velcurve}, we
describe the method of determining radial velocities and discuss the
resulting radial-velocity curves.  Sect.~\ref{sec:iue} is an
intermezzo, dedicated to the reanalysis of IUE spectra, a subset of
which was already used in Paper~II in the determination of the mass;
we compare our results to the ones found by Stickland et al.\
(\cite{SLR97}) who analysed the same set of IUE data, and explain the
origin of the difference between our results.  In
Sect.~\ref{sec:allvel}, we combine all velocity curves.  We also
investigate systematic deviations with orbital phase, finding that,
unfortunately, these are present and that they limit our final
accuracy.  In Sect.~\ref{sec:mass}, we present our results for the
mass of the neutron star and in Sect.~\ref{sec:discussion} we
discuss how these results influence the current status of our
knowledge on the internal structure of neutron stars.

\section{Data reduction and analysis}\label{sec:reduction}

We observed HD~77581, the bright ($V=6.9$~mag) B-supergiant companion
to \VelaX1, in the period October 15, 1995, until July 15, 1996,
using the 1.4-m Coud\'{e} Auxiliary Telescope (CAT) and Coud\'{e}
Echelle Spectrograph (CES) at the European Southern Observatory (ESO),
La Silla, Chile.  The observations were carried out by the night
assistant of the CAT (in La Silla or, more often, remotely from the
ESO Headquarters in Garching, Germany) or by the astronomer present at
the telescope.

Every night one spectrum (in a few cases two) of HD~77581 was
obtained, as well as a few bias frames, two internal lamp flat fields
and a Th-Ar wavelength calibration frame.  Exposure times range from
20 to 30 minutes and the typical signal-to-noise ratio of the
extracted spectra is $\ga\!300$ per $5~\kms$ resolution element.  A
fraction of the obtained spectra (47 out of 230) could not be used due
to various practical problems (e.g., wavelength region covered by the
stellar spectrum was not identical to that in the
wavelength-calibration frame, no calibration frame available, wrong
exposure level science frame, etc.).

The CES spectrograph has two ``paths'', one optimized for the blue and
one for the red part of the optical spectrum.  The CES long camera is
equiped with one of two CCD detectors, CCD\#34 ($2048\times2048$
pixels) or CCD\#38 ($2720\times520$ pixels, and thus a larger
wavelength coverage).  In a given night, one has to select the red or
the blue path, as well as the CCD.  In order to minimize the impact of
our programme on the observing run carried out during the remaining
part of the night, we had to observe with the setting selected by the
observer of that night.  For this reason, we selected two wavelength
regions, one in the blue (4077--4127~\AA) and one in the red path of
the spectrograph (5670--5720~\AA).  A total of 104 usable spectra were
taken in the red and 79 in the blue wavelength region. A log of the
observations is given in Tables~\ref{tab:blue} and \ref{tab:red} in
the Appendix.

One of the main problems we were faced with during the analysis is the
vignetting in the science frames, which varies from night to night,
depending on the position of the telescope.  It is caused by the small
size of the field lens of the long camera (which was not designed for
large CCDs), and the variations are due to the large distance between
spectrograph and telescope. The CAT telescope is positioned in a dome
just next to the dome of the 3.6m telescope, while the CES
spectrograph is located inside the 3.6m dome. The light coming from
one of the Nasmyth foci of the CAT goes through a pipe connecting the
two domes, into the predisperser area of the spectrograph, bridging a
distance of about 15~m.  For this reason, it is very difficult
(especially in view of the regular earthquakes) to align the telescope
and spectrograph beam.  Dome flats provide a better way to correct for
this misalignment than internal flats; unfortunately, in practice it
was not feasible to obtain dome flats for our programme.  The vignetting
is more severe at longer wavelengths.  The correction for the
vignetting, though hampered by the relatively small wavelength region
covered in one spectrum, has been carried out during the normalisation
procedure.

Below, we describe the extraction, calibration, and
normalisation of the spectra.

\subsection{Bias subtraction}\label{sec:bias}

The electronic bias distribution is uniform over both chips, therefore
in the bias correction (per night) we subtracted a number, derived
either from the mean of the bias frames taken in that night (for the
spectra taken from October 95 until March 96), or from the overscan of
the science frame itself (for the spectra taken from March 96 until
July 96).

\subsection{Flatfielding}\label{sec:flat}

In order to apply the flatfield correction, we constructed one
``master'' flatfield frame for each of the two CCDs and the two
wavelength regions, i.e., four frames in total. The master flatfield
consists of the average of all available flatfields in a given
wavelength region and taken with a given CCD, after inspection for
saturation or under-exposure and a consistency check of their spectral
parameters (central wavelength, resolution power, pre-disperser exit
slit width). The individual flatfields were normalised to unity before
averaging.  Since the position of the spectrum on the CCD was not
always the same during the extended period of our observing campaign,
the master flatfield does not have a uniform exposure level.  We
scaled the flatfield to unity by dividing each row of the detector by
its mean value.  The night-to-night changes in instrument setting are
small and have no significant effect on the flatfield structure (e.g.,
we were not confronted with strong detector fringes or ghost images).
 
\begin{figure}
\centerline{\psfig{figure=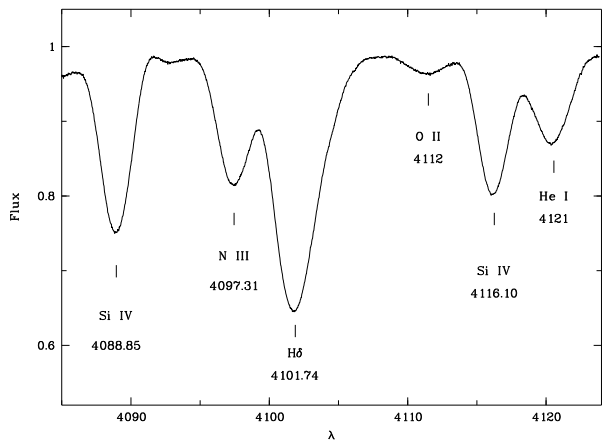,height=9cm,angle=-90}}
\caption[]{The average blue CAT/CES spectrum centered at H$\delta$
4102~\AA, with line identifications. The wavelength scale is in
\AA ngstr\o m.}
\label{fig:blueav}
\end{figure} 

\subsection{Optimal extraction}\label{sec:extraction}

We used an optimal extraction algorithm developed by Van Kerkwijk
(\cite{vKer93}), based on the method of Horne (\cite{Hor86}), which
includes corrections for small non-alignments of the spectrum with the
dispersion direction of the chip, and detects and filters for
cosmic-ray events. Some bad columns on the detectors have been
corrected for by interpolation (in pixel space) over the gaps in the
extracted spectrum.
 
In the red wavelength region some atmospheric water vapour lines are
present. As they would cause problems in the velocity determination we
removed them by linear interpolation.

\subsection{Wavelength calibration}\label{sec:wavecal}

For the wavelength calibration we use the Th-Ar calibration frames,
fitting the line positions with a second-order dispersion relation.  The
typical root-mean-square residuals are 0.001~\AA.  The dispersion is
0.018~\AA\ per pixel for the blue region and 0.024~\AA\ per pixel for
the red region.  There are about five pixels per resolution element.

\subsection{Barycentric correction and orbital ephemeris}\label{sec:timing}

We corrected for the Earth's orbital and rotational motion by
calculating the component of the Earth's velocity in the direction of
HD~77581 at the mid-time of the observation and subtracting it from
the velocity measured from the determined Doppler shift of the line
(see Sect.~4).  The orbital phase was calculated using the orbital
parameters based on the BATSE data (Bildsten et al. \cite{BCC+97});
these are listed in Table~\ref{tab:batse}.

\begin{figure}
\centerline{\psfig{figure=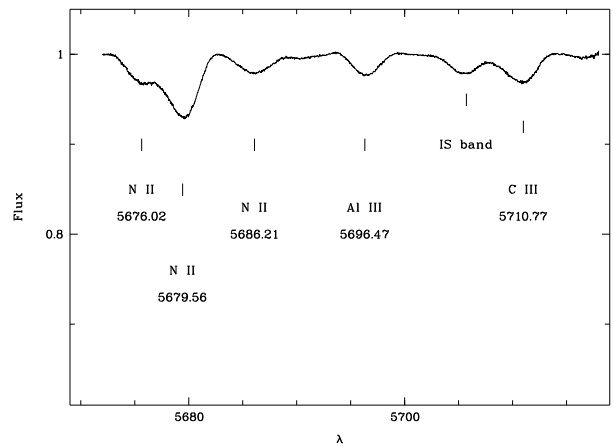,height=9cm,angle=-90}}
\caption[]{The average red CAT/CES spectrum centered at 5695~\AA,
which includes several lines of a \ion{N}{2} multiplet. Note the
difference in strength between the lines present in the blue and red
spectra.} 
\label{fig:redav}
\end{figure} 

\subsection{Normalisation of the spectra}\label{sec:norm}

Normalising the spectra was not a straightforward procedure, the main
reason for this being the (time-dependent) vignetting (see above)
which affects all spectra.  We explored different methods to normalise
the blue and the red spectra, taking into account: (i) the use of the
same continuum regions (after shifting the spectra approximately to
the reference frame of the star) for each spectrum and (ii) the
assumption that the shape of the spectral continuum can be described
by a polynomial of order four or lower.

The main problem we were confronted with was the lack of clean
continuum regions due to the small wavelength region ($\sim 50$~\AA)
covered in each spectrum. Also, the intrinsic (i.e. non-orbital)
variability of the stellar spectrum sometimes causes the lines to
extend into the spectral continuum regions. 

Below we describe the normalisation methods we preferred after
testing different approaches.  We note that we repeated the full
analysis (including the cross-correlation procedure) for different
normalisation methods, but did not find significantly different
results.

\subsubsection{Blue spectra}\label{sec:normblue}

The central part of the blue spectrum (Fig.~\ref{fig:blueav}), i.e.,
at the blend of \ion{N}{3}~4097~\AA\ and H$\delta$, is not affected
by vignetting, so the shape of the continuum is well reproduced by a
linear function defined by the continuum regions left and right of the
blend. The edges of the spectrum were normalised using two
second-order polynomials, one at each side of the spectrum, which were
forced to have the same slope as the linear function at the point
where they meet. The normalisation regions are given in
Table~\ref{tab:norm}.

\begin{table}
\caption[]{The wavelength regions used to normalise the spectra, for
the blue and red wavelength region and CCD\#34 and 38, respectively.
Before normalisation, the selected continuum regions are shifted by an
amount corresponding to the (predicted) Doppler shift for the orbital
phase at the time of observation.}
\label{tab:norm}
\begin{tabular}{ll}
\hline\relax\\[-2ex]
CCD\#38 &CCD\#34\\
\hline\relax\\[-2ex]
\multicolumn{2}{l}{\em Blue}\\
4080.45--4082.10& 4086.04--4086.50\\
4091.25--4091.77& 4091.50--4092.50\\
4107.10--4108.00& 4094.35--4094.95\\
4122.62--4124.05& 4108.10--4108.50\\
                & 4113.20--4113.70\\
                & 4118.22--4118.77\\[.3ex]
\multicolumn{2}{l}{\em Red}\\
5672.30--5672.90& 5673.40--5673.55\\
5682.82--5683.38& 5682.82--5683.38\\
5693.00--5693.66& 5693.00--5693.66\\
5699.98--5702.58& 5699.61--5700.38\\
5714.25--5716.00& 5713.04--5714.04\\[.3ex]
\hline
\end{tabular}
\end{table}

\subsubsection{Red spectra}\label{sec:normred}

The effect of vignetting is more pronounced in the red wavelength
region. Furthermore, the intrinsic variability of the line profiles is
much larger compared to that in the blue spectrum, the spectral lines
are closer to each other, and sometimes there appear to be (variable)
P~Cygni emission features at the blue side of the strongest
\ion{N}{2} lines.  There are no extended continuum regions which are
completely devoid from spectral features.  The regions listed in
Table~\ref{tab:norm} are those we believe are least affected by
spectral lines.

In order to reduce the effect of the vignetting (which otherwise would
force us to use a high-order polynomial to fit reasonably the shape of
the continuum) we divided all the spectra by one spectrum (r8005r)
which appears to be only moderately affected by vignetting.  The
residuals were fitted with a third-order polynomial.  The latter fits
should approximately reproduce the shape of the continuum and
eliminate most of the vignetting.  Then, we multiplied the original
spectra with this fit.  In this way, the spectra were ``ironed''.
Finally, we fitted a fourth-order polynomial to the ironed spectra,
using the normalisation regions listed in Table~\ref{tab:norm}.

\section{Description of the spectra}\label{sec:spectra}

In Figs.~\ref{fig:blueav} and \ref{fig:redav} we show the average of
the blue and of the red spectra, respectively, after having taken into
account the difference in orbital motion.  The blue spectrum includes
stronger lines than the red spectrum, which makes the blue spectrum
better suited to measure the star's radial velocity.

An overview of the blue and red spectra is presented in
Figs.~\ref{fig:blue38} and \ref{fig:red38}, respectively.  The spectra
are ordered according to orbital phase; the corresponding observing
times can be found in Tables~\ref{tab:blue} and \ref{tab:red} in the
Appendix. The few spectra obtained with CCD\#34 (instead of CCD\#38)
are shown in Figs.~\ref{fig:blue34} and \ref{fig:red34}. The
wavelength scale of the spectra is with respect to the stellar rest
frame, i.e., the orbital motion has been taken out to facilitate
comparison.  The figures show that the profiles of the lines in the
red spectra are much more variable than those of the lines in the blue
spectra.  In a given spectrum, the different lines seem to vary mostly
in concert.  In some cases the variations are dramatic, especially in
the red spectra, where the depths of the photospheric lines are
sometimes strongly reduced.  The variations do not show an obvious
dependence on orbital phase.  In order to have a better view on
possible orbital-phase related line-profile variations, we binned the
spectra in nine phase bins (nine bins is chosen since the orbital
period is very close to nine days).  The results are shown in
Figs.~\ref{fig:bluephaseav} and \ref{fig:redphaseav}.
                           
\subsection{The blue spectra centered on
H$\delta$~4102~\AA}\label{sec:bluespec}  

The blue spectra cover H$\delta$ and two strong \ion{Si}{4} lines, at
4088.85 and 4116.10\,\AA, as well as weaker lines of \ion{N}{3} (at
4097.31\,\AA), \ion{O}{2} (4112\,\AA) and \ion{He}{1} (4121\,\AA).
Furthermore, there are two small features which we could not identify;
these lines are also seen in the average spectrum in Paper~II.  These
are located at 4084.88\,\AA\ just shortward of the \ion{Si}{4}
4088.85\,\AA\ line and at 4092.96\,\AA\ at the opposite side of the
\ion{Si}{4} line.

In general, the lines show strikingly similar behaviour.  For the two
\ion{Si}{4} lines this is expected, since they arise from the
same multiplet.  But the \ion{N}{3} line at 4097.31\,\AA, while
blended with the H$\delta$ line, generally behaves similarly as well.
See, e.g., the blue spectrum at phase 0.322 in
Fig.~\ref{fig:blue38}, where the profiles of the \ion{Si}{4} lines
and the \ion{N}{3} line show an identical deformation of line shape.
It is less easy to judge the case for the \ion{O}{2} line, which is
rather weak, and the \ion{He}{1} line, which is not completely covered
by our spectra (hence we did not include the latter in our analysis).

\begin{figure*}
\vspace{10cm}
\caption[]{The blue spectra obtained with CCD\#38 ordered according to
orbital phase.  The Doppler shift due to the orbital motion has been
taken out.  Note the variability of the line profiles, and the
strong correlation in shape between the different lines.  The
H$\delta$ line shows stronger variations in the blue wing when the
X-ray source passes through the line of sight ($\phi=0.5$).  These
variations are likely due to the presence of a photo-ionization wake
in the B-supergiant's wind.}
\label{fig:blue38}
\end{figure*} 

\begin{figure}
\centerline{\psfig{figure=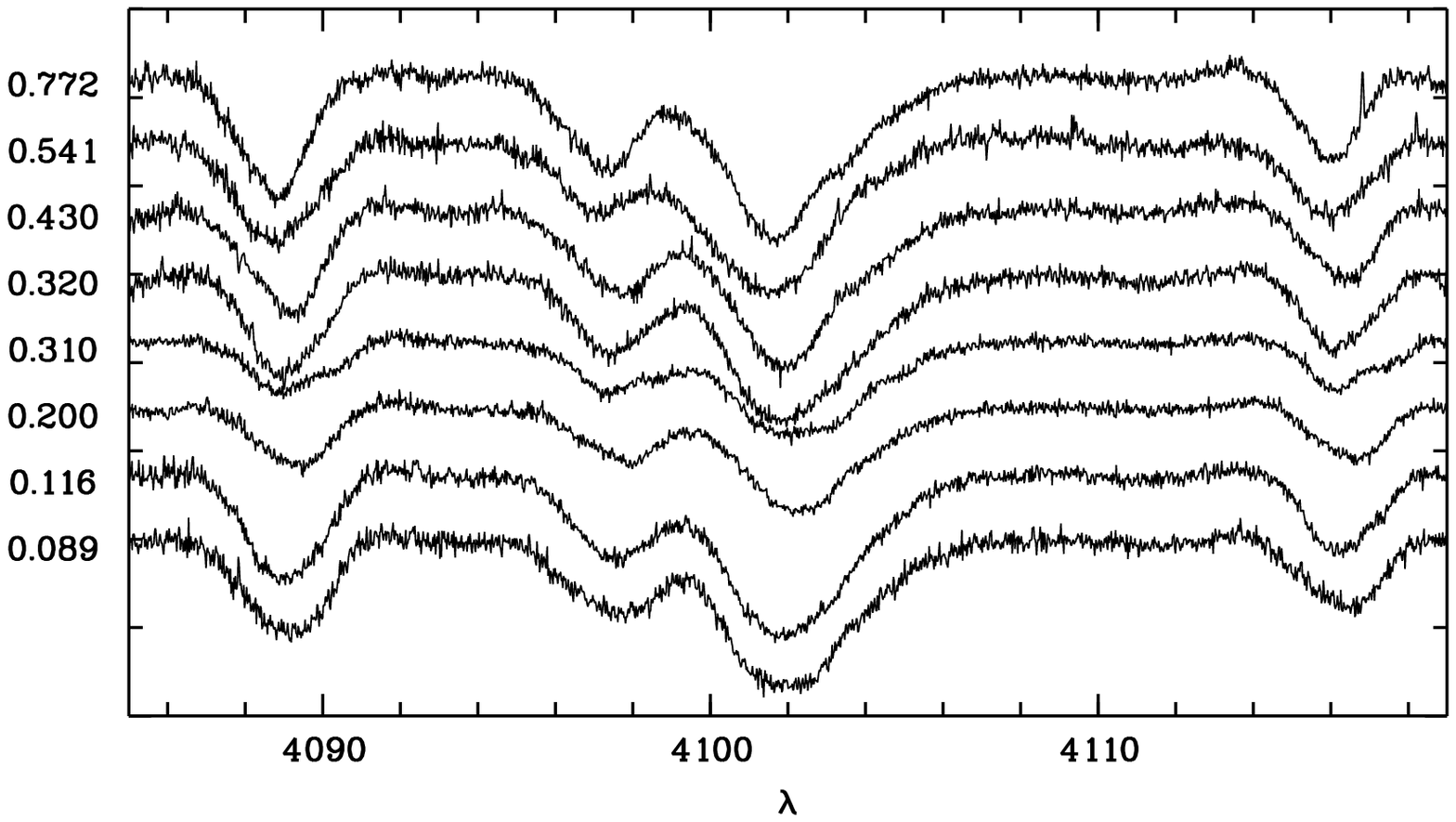,width=8.5cm,clip=}}
\caption[]{The blue spectra obtained with CCD\#34, ordered according
to orbital phase.}
\label{fig:blue34}
\end{figure} 

\begin{figure}
\centerline{\psfig{figure=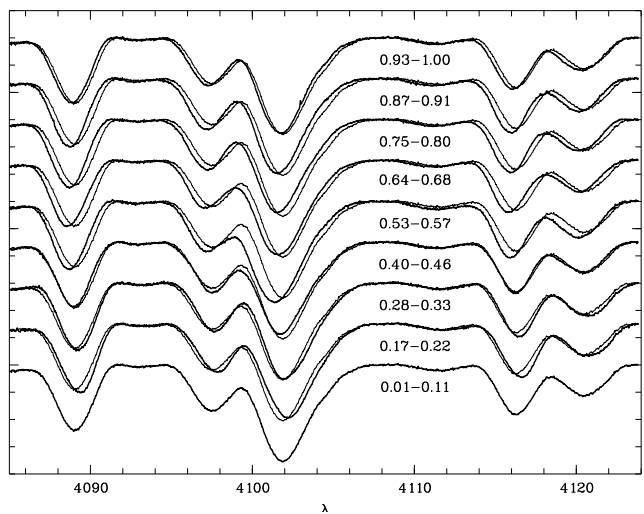,angle=-90,width=8.5cm,clip=}}
\caption[]{Mean blue spectra, obtained after averaging over nine
different phase bins. The dotted line represents the mean profile at
orbital phase 0.01-0.11 and is plotted for comparison.}
\label{fig:bluephaseav}
\end{figure} 

Also in the phase-binned spectra, the lines appear mostly to vary in
concert.  The profiles close to X-ray eclipse (in phase intervals
0.01--0.11 and 0.95--1.00) are rather symmetric, while the mean
profiles in phase intervals 0.17--0.33 and 0.75--0.80 are shifted to
the red, and those from 0.53--0.64 appear asymmetric to the blue.  The
variations are quite a bit smaller, however, than the large internal
spread within each phase interval.

The only line showing deviant behaviour is H$\delta$.  During orbital
phases 0.45--0.65, i.e., when the neutron star passes in front of the
supergiant companion, the line profile is strongly distorted,
especially the blue wing, where often a weak and blue-shifted
absorption component can be observed, which can have such a high,
blue-shifted velocity that it blends with the \ion{N}{3} line.  At
other phases, the line profile behaviour is more similar to what is
seen for the \ion{Si}{4} and \ion{N}{3} lines.

\begin{figure*}
\vspace{10cm}
\caption[]{The red spectra (CCD\#38) ordered according to orbital
phase.  The lines of the \ion{N}{2} multiplet show strong intrinsic
variability, apparently unrelated to orbital phase. In the most
extreme cases, the lines almost disappear.}
\label{fig:red38}
\end{figure*}

\begin{figure}
\centerline{\psfig{figure=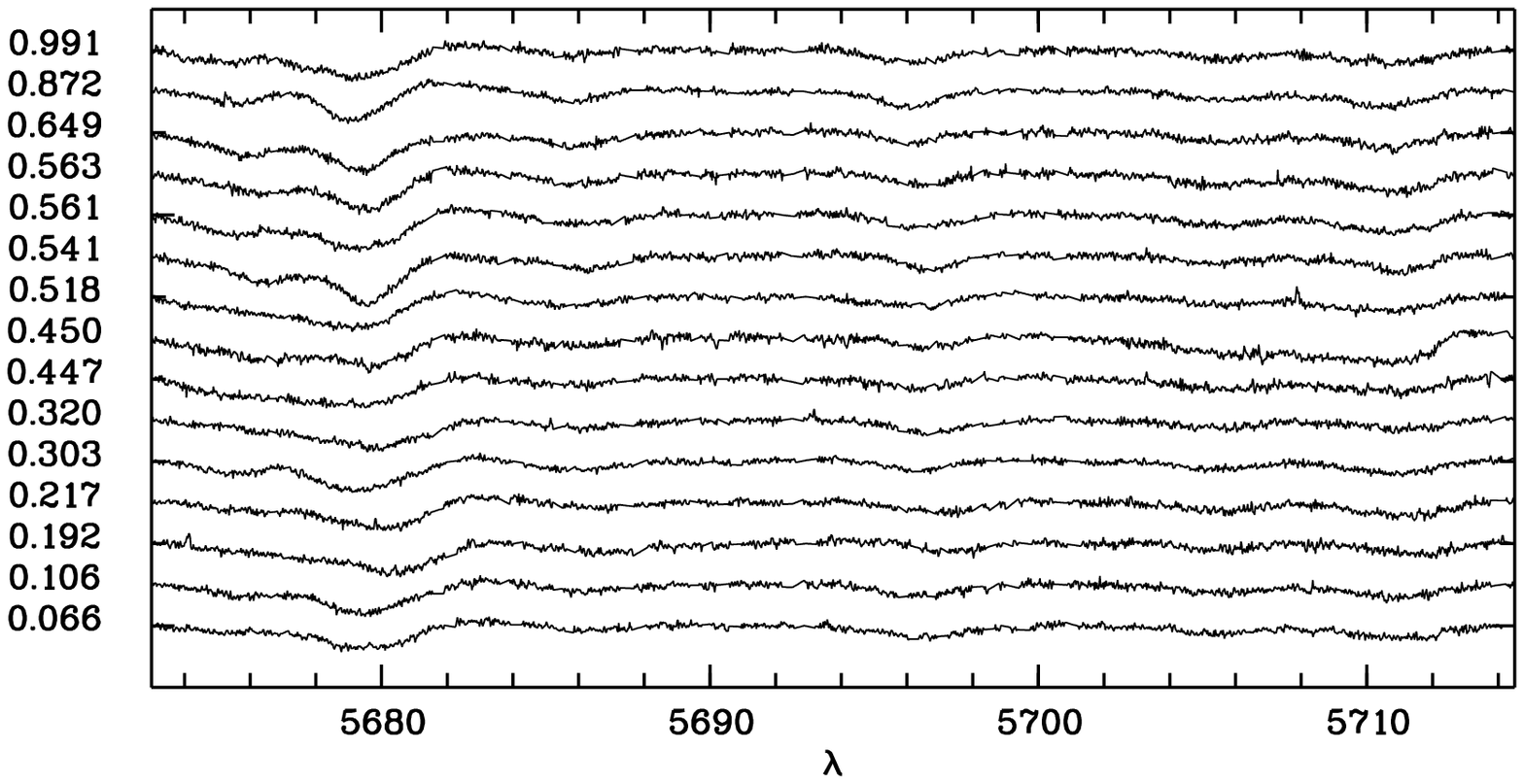,width=8.5cm,clip=}}
\caption[]{The red spectra obtained with CCD\#34, ordered according to
orbital phase.}
\label{fig:red34}
\end{figure} 

\begin{figure}
\centerline{\psfig{figure=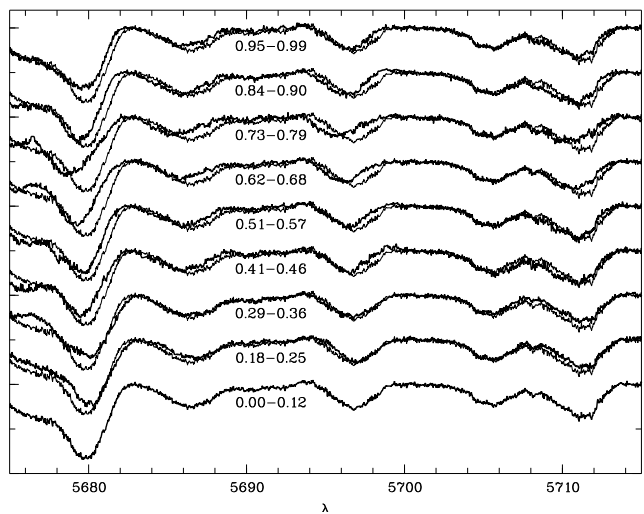,angle=-90,width=8.5cm,clip=}}
\caption[]{Mean red spectra, obtained after averaging over nine
different phase bins. The dotted line represents the mean profile at
orbital phase 0.01-0.11 and is plotted for comparison.}
\label{fig:redphaseav}
\end{figure} 

In Fig.~\ref{fig:bluephaseav}, the enhanced blue absorption in
H$\delta$ can be seen clearly in the mean profiles of phase intervals
0.40--0.46 and 0.53--0.57.  This distortion of the H$\delta$ line is
very similar to that seen in the H$\beta$ and \ion{He}{1} 4471~\AA\
line of HD~77581 (Kaper et al.\ \cite{KHZ94}), and is most likely
caused by the perturbation of the B-supergiant wind due to the
presence of the X-ray source (see Sect.~\ref{sec:ionisationwake}).

\subsection{The red spectra centered at 5695\,\AA}\label{sec:redspec}

The red spectra cover lines which are much weaker than those in the
blue, and which, as is clear from the spectra, also show much stronger
intrinsic perturbations of their profiles.  The lines covered include
\ion{N}{2} lines at 5676.02, 5679.56, and 5686.21\,\AA, an \ion{Al}{3}
line at 5696.47\,\AA, and an \ion{C}{3} line at 5710.77\,\AA.
Furthermore, the spectra contain a diffuse interstellar band at
5705~\AA.  Its constant profile demonstrates the stability of the
instrument, but at the same time, since it is contaminating the
\ion{C}{3} line it hinders us from using the latter line to measure
velocities.  Finally, between the \ion{N}{2} and the \ion{Al}{3} line,
at $\lambda$5691.10~\AA, there is a very shallow variable absorption
feature which we are not able to identify.

The \ion{N}{2} lines belong to the same multiplet.  All three lines
show dramatic intrinsic variations, with, like for the blue spectra,
little evidence for a correlation with orbital phase.  On occasion,
the distinction between the two blended \ion{N}{2} lines disappears
completely (see, e.g., phases 0.184 and 0.511), or an additional
feature occurs (phases 0.337 or 0.458), in an apparently irregular
manner.  The strongest \ion{N}{2} line sometimes appears to develop a
red (P-Cygni) emission bump.  If real, this would be an indication
that the strongest \ion{N}{2} line is formed in the outer expanding
layers of the atmosphere.  Note that the \ion{N}{2} lines are often a
useful diagnostic in studying non-radial pulsations in B-type stars
(D.R.\ Gies, private communication).  Fig.~\ref{fig:redphaseav}
indicates that, on average, at orbital phases close to X-ray eclipse
(0.95--1, 0--0.25) the red spectral lines are deepest.

In general, the \ion{Al}{3} and \ion{C}{3} line profiles are similar
to those of the \ion{N}{2} lines.  See, e.g., the spectra at orbital
phases 0.868 and 0.872, where all the line profiles (except that of
the interstellar band, of course) are distorted towards longer and
shorter wavelengths, respectively.  Again, at several orbital phases
(e.g., 0.01) features appear which could be interpreted as being due
to incipient emission.  These can be present either on the left wing
(\ion{Al}{3} line, e.g., phases 0.01, 0.296, and 0.675), or on the
right wing (\ion{Al}{3} line, phase 0.448).  In view of the difficulty
of defining the continuum level, it is hard to say whether these are
indeed emission features or whether there is a weak absorption feature
right next to it.

Sometimes (see, e.g., the spectra obtained at phases 0.653 and 0.78)
the depth of the lines decreases so much that the \ion{N}{2} blend
becomes very shallow, whereas the weaker \ion{N}{2} and \ion{Al}{3}
lines almost disappear.

\section{Radial-velocity determination}\label{sec:radvel}

\begin{figure*}
\centerline{\vbox{
\hbox to\hsize{\hfill
\psfig{figure=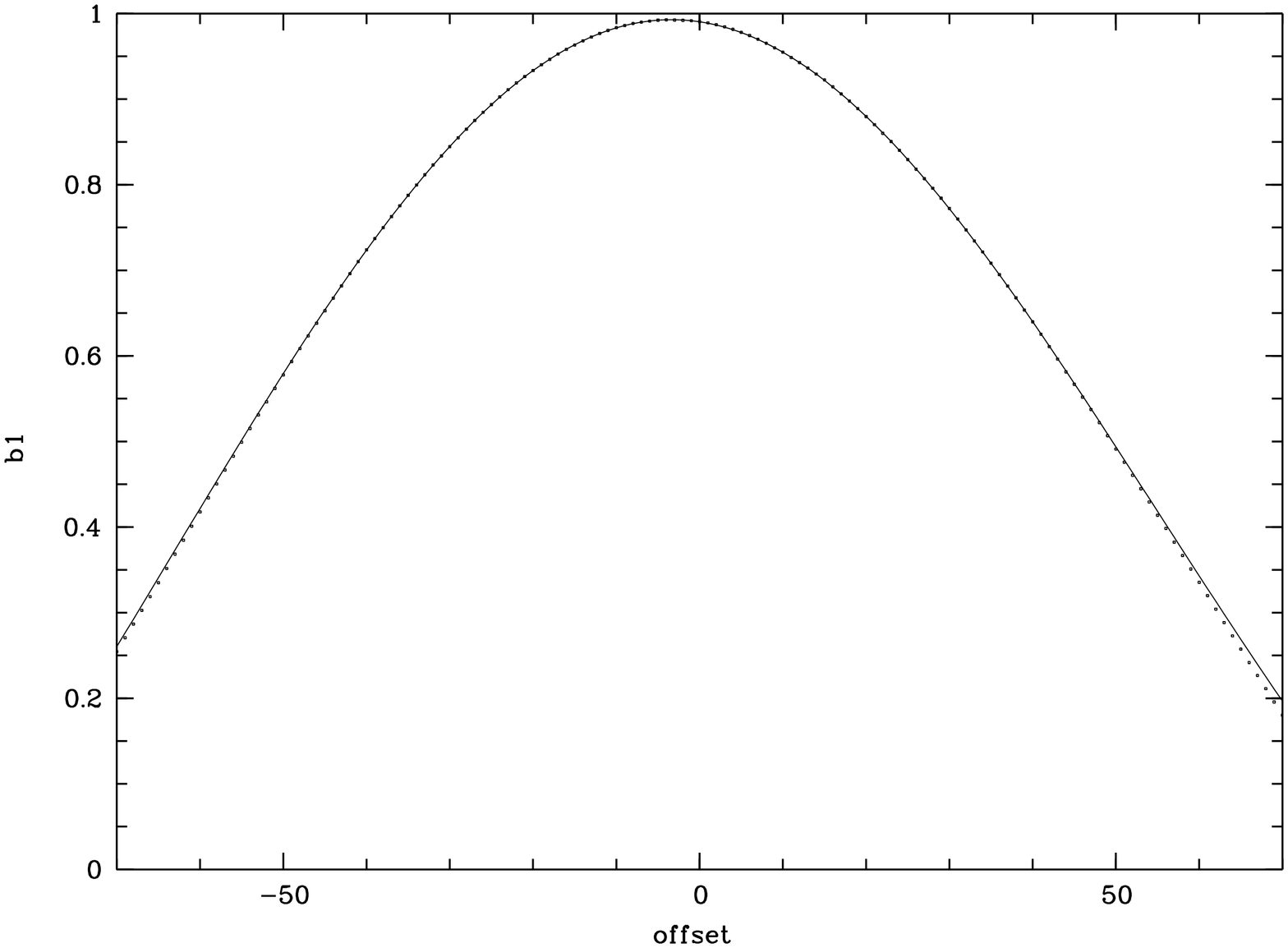,height=8cm,width=6cm,angle=-90}\hfill
\psfig{figure=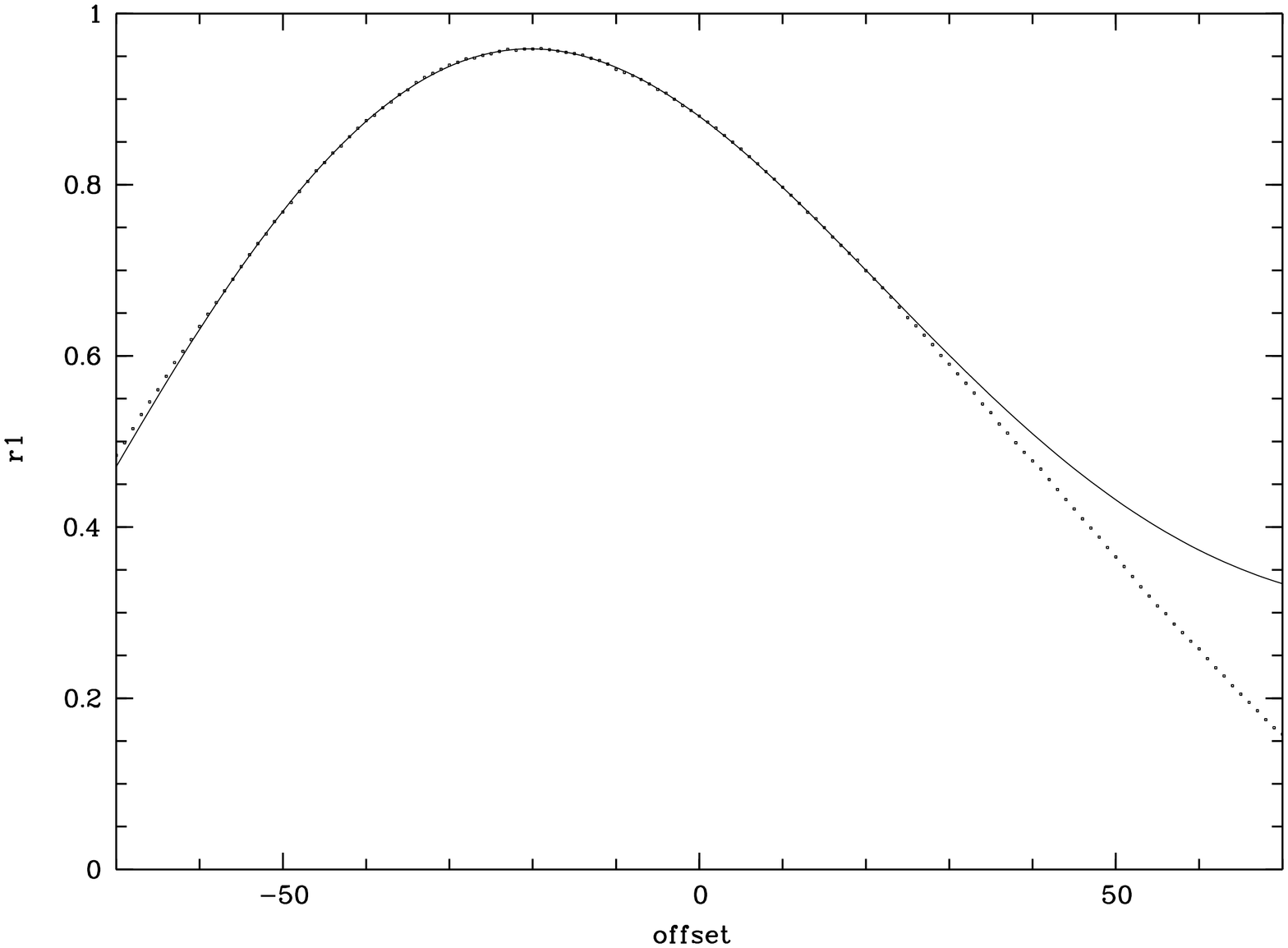,height=8cm,width=6cm,angle=-90}\hfill}
\hbox to\hsize{\hfill
\psfig{figure=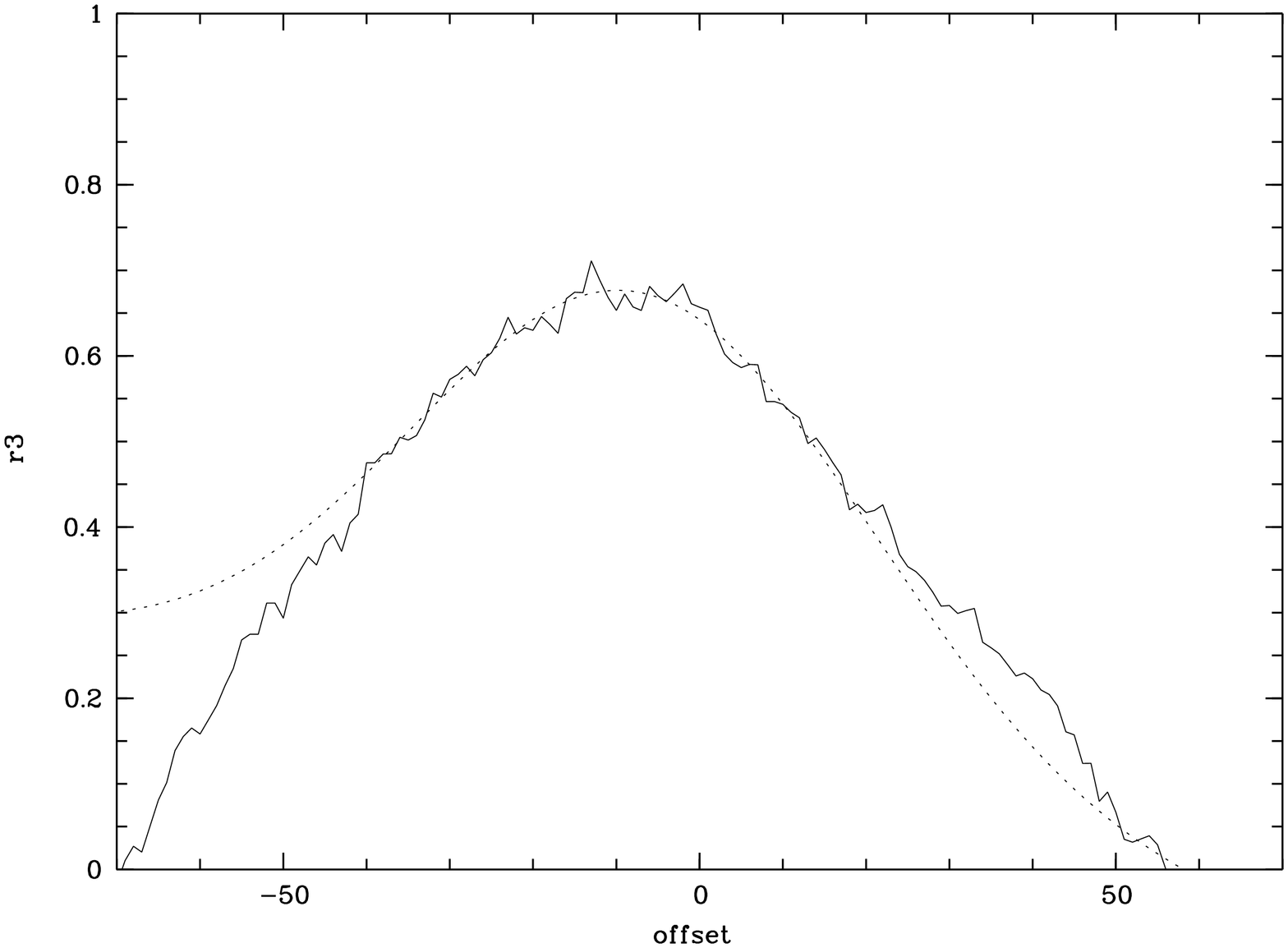,height=8cm,width=6cm,angle=-90}\hfill
\psfig{figure=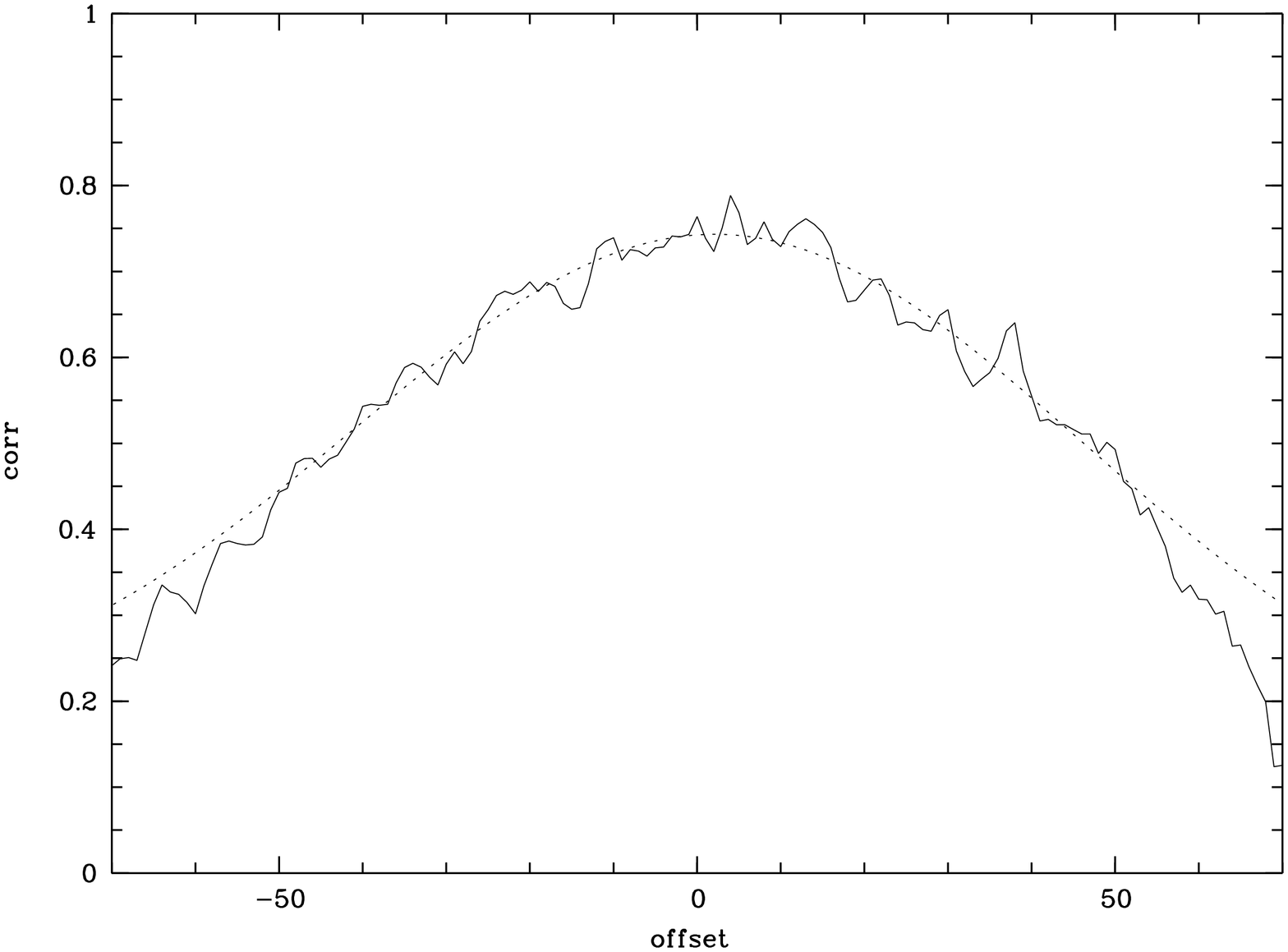,height=8cm,width=6cm,angle=-90}\hfill}}}
\caption[]{A typical cross-correlation function and the fit applied to
the area near the maximum for \ion{Si}{4} 4088.85 \AA\ ({\em upper
left}), \ion{N}{2} 5676/5679~\AA\ blend ({\em upper right}),
\ion{Al}{3} 5696.47~\AA\ ({\em lower left}) and the interstellar band
({\rm lower right}).  The offset is measured in~\kms.}
\label{fig:ccf}
\end{figure*} 

We determined the radial velocity of HD~77581 by means of
cross-correlation.  We used the same algorithm as in Paper~II (see also
Van Kerkwijk \cite{vKer93}) where each spectrum is cross-correlated
with all other spectra in the data set.  We decided to use all the
available spectra as independent reference spectra, because of the
lack of a template spectrum of a bright star with identical spectral
type.  We did not cross-correlate against the average spectrum to
avoid auto-correlation effects which lead to a systematic decrease of
the amplitude of the radial-velocity curve.  For details on the
cross-correlation method we refer to Paper~II and Van Kerkwijk
(\cite{vKer93}).

We first transformed the spectra to a logarithmic wavelength scale so
that a Doppler shift is a linear displacement along the spectrum.  The
top part of the cross-correlation function (CCF) was fitted with an
analytic function (a Gaussian plus a linear function) whose maximum
determines the velocity shift of the spectrum.  Obviously, the
relevant part of the cross-correlation function is the area close to
the maximum; we fit only the upper 30 to 40\% of the CCF, depending on
its shape and smoothness (Fig.~\ref{fig:ccf}).

We cross-correlated all the lines separately as well as combined,
using the wavelength regions listed in Table~\ref{tab:ccregions}.  The
measured radial velocities are listed in Table~\ref{tab:blue} (blue
wavelength region) and Table~\ref{tab:red} (red wavelength region) in
the Appendix.  The typical intrinsic, formal accuracies are $0.7~\kms$
for the blue spectra and $1.5~\kms$ for the red spectra.

\begin{table}
\caption[]{Cross-correlation regions.}
\label{tab:ccregions}
\begin{tabular}{ll}
\hline\relax\\[-2ex]
CCD\#38 &CCD\#34\\
\hline\relax\\[-2ex]
\multicolumn{2}{l}{\em Blue}\\
4084.40--4092.45& 4085.00--4093.30\\
4092.50--4099.04& 4093.30--4099.40\\
4108.30--4118.12& 4110.13--4119.12\\[.3ex]
\multicolumn{2}{l}{\em Red}\\
5669.76--5682.98& 5673.20--5682.98\\      
5682.98--5693.78& 5682.98--5693.78\\      
5693.78--5701.91& 5693.78--5701.91\\[.3ex]
\hline
\end{tabular}
\end{table}

The cross-correlation method worked satisfactorily for the blue
spectra.  The CCF is very smooth and the fit well defined.  This
results in an accurate determination of the velocities.  In
Fig.~\ref{fig:ccf}, we show a typical CCF, derived from the
cross-correlation of the \ion{Si}{4} line, as well as the fit.

In the red spectra, the CCF is much less well-defined, since the
absorption features are much weaker.  We partly overcame this by
allowing the fit to the CCF to cover a larger area around the peak so
that the determination of the peak position is not too much affected
by the noise.  Fortunately, the \ion{N}{2} blend (at 5676.02 and
5679.56\,\AA), despite its strong intrinsic variations, is
sufficiently deep to produce a reasonable CCF.  In Fig.~\ref{fig:ccf},
we show a typical CCF of the \ion{N}{2} blend and of the
\ion{Al}{3} line.

\begin{figure}
\centerline{\psfig{figure=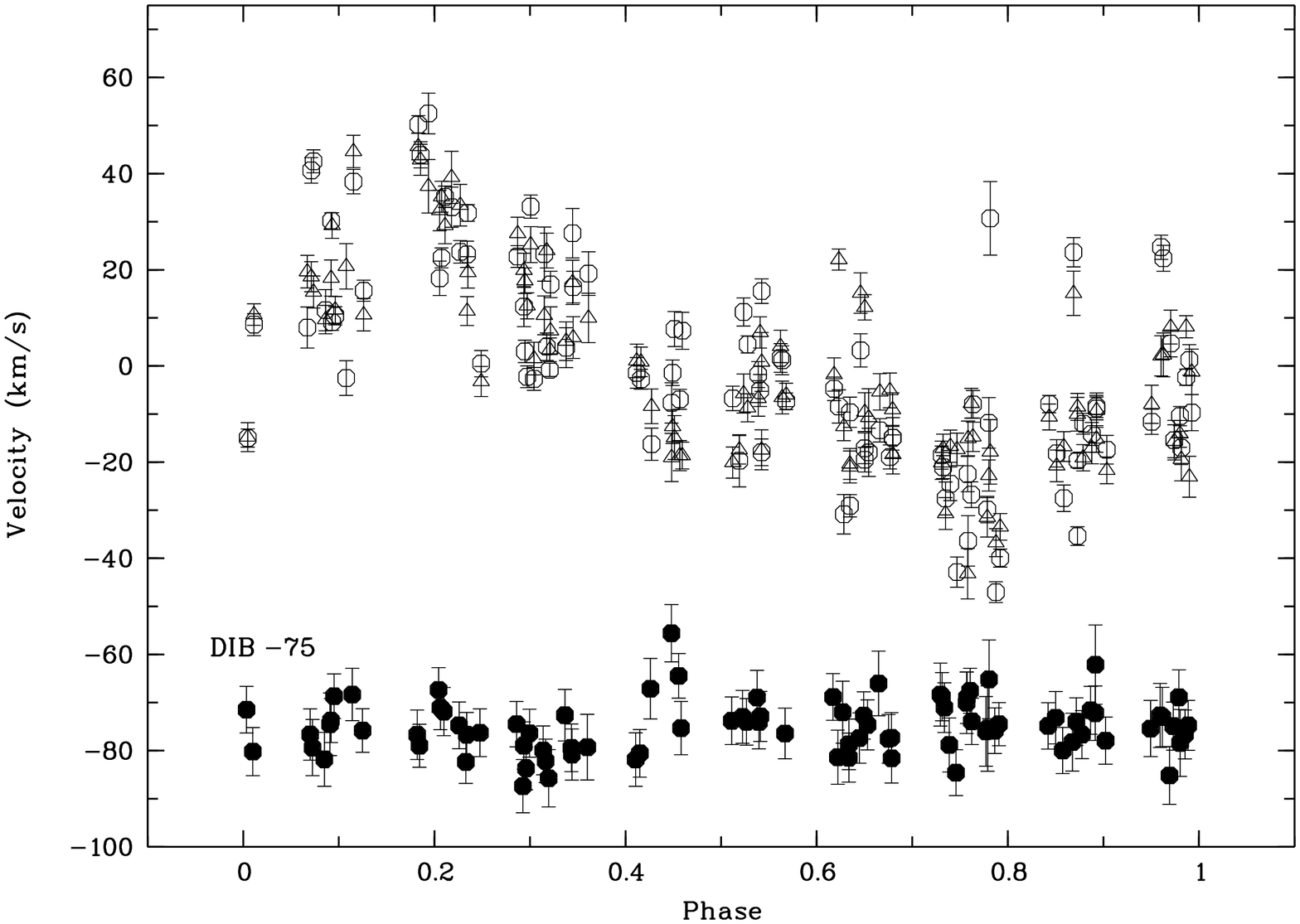,angle=-90,width=8.8cm}}
\caption[]{The radial velocity of the diffuse interstellar band (DIB)
as a function of orbital phase (filled circles, shifted by
$-75{\rm\,km\,s^{-1}}$). For comparison, the radial velocities are
shown for the \ion{N}{2} 5686\,\AA\ (triangles) and \ion{Al}{3} 
(open circles) regions; these lines have comparable depth in the red
spectra.  As expected, the velocity of the DIB is consistent with a
constant value.}
\label{fig:isvel}
\end{figure}

The presence of the interstellar line at 5705\,\AA\ allows us to check
for trends, e.g., due to instrumental effects, in the derived radial
velocities.  In the bottom-right panel in Fig.~\ref{fig:ccf} we show
the CCF for the diffuse interstellar band.  It is rather wiggly, since
the band is quite weak.  Fig.~\ref{fig:isvel} displays the radial
velocities derived from the band (filled circles, displaced by
$-75{\rm\,km\,s^{-1}}$).  For comparison, the radial-velocity curves
of the \ion{N}{2} 5686\,\AA\ and \ion{Al}{3} line regions are shown.
The radial velocity of the DIB is consistent with a constant value
($\chi^2_{\rm{}red}=1.1$).  A slight phase dependence might be
introduced due to the neighbouring \ion{C}{3} line, which moves close
to the DIB around phase 0.75, but this does not become apparent in
Fig.~\ref{fig:isvel}.  Nevertheless, to be on the safe side, we did
not use the radial velocities measured from the \ion{C}{3} line in our
solutions.

From the above, we conclude that our error estimates are reasonable
for the interstellar band.  This feature does not vary in shape, and
hence it satisfies the assumption underlying cross-correlation, viz.,
that the two spectra being correlated differ only in their radial
velocity.   This assumption will not hold for the stellar lines, given
the large variations seen in the line profiles.  Therefore, the error
estimates become somewhat ill-defined.  Nevertheless, if all lines
varied truly in concert, one would expect the inferred velocities to
be the same within the errors.  

\begin{figure*}
\centerline{\hbox{
\psfig{figure=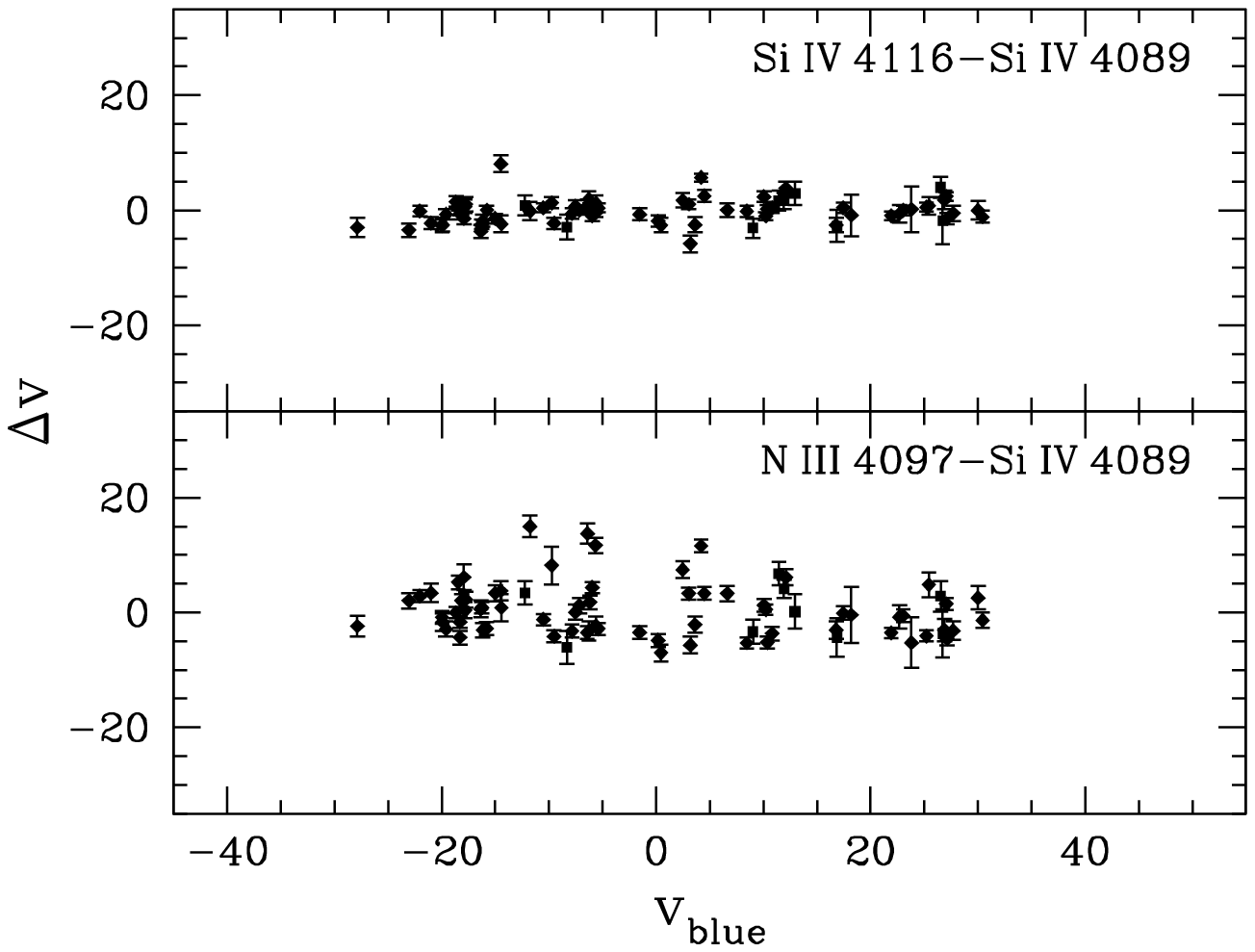,width=8cm} \hspace{1cm}
\psfig{figure=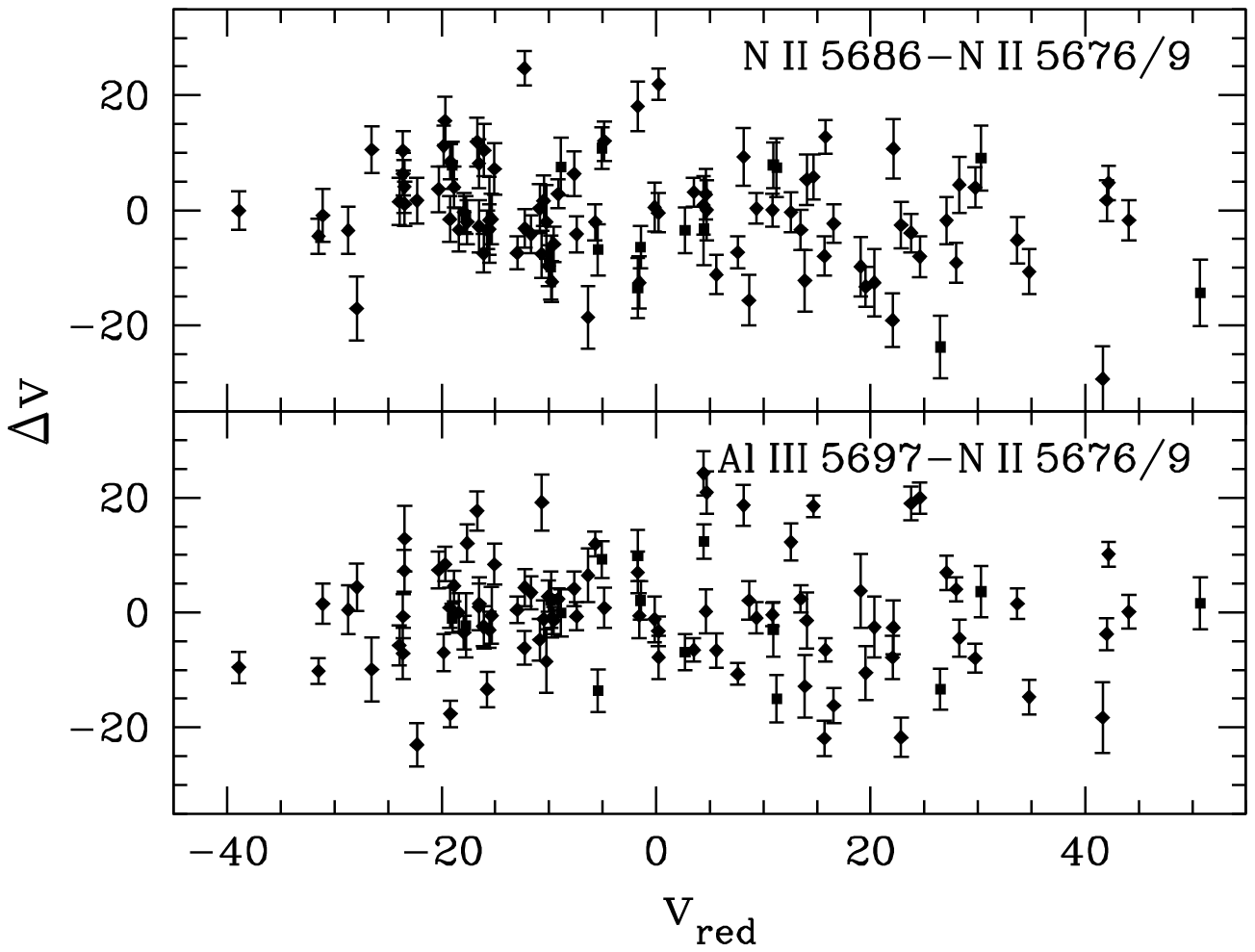,width=8cm}}}
\caption[]{Comparison of the radial velocities derived from different
spectral lines. The difference in radial velocity is shown with
respect to the radial velocity derived for one of the lines. The label
at the top of each figure indicates which lines are involved. The left
and right figures relate to the blue and red spectral domain,
respectively.}
\label{fig:velcomp}
\end{figure*}

In Fig.~\ref{fig:velcomp}, we compare the radial velocities derived
for the different lines.  For the two regions centred on the
\ion{Si}{4} lines, we find a good correspondence.  However, while
there are no systematic deviations, the velocities inferred for the
lines are not consistent with each other within the formal
uncertainties (root-mean-square difference of 2.1~\kms;
$\chi^2_{\rm{}red}=3.5$), indicating that there are small differences.
We do not believe that this could be due to differences in the
behaviour of the two \ion{Si}{4} lines, which, as mentioned, arise
from the same multiplet.  Instead, it might be due to differences in
other lines in the two regions; e.g., the 4116~\AA\ region is
influenced by the wing of the \ion{He}{1}~4121~\AA\ line.

Indeed, the velocities inferred from the \ion{N}{3} region show larger
differences with those inferred from the \ion{Si}{4} lines (rms of
4.6~\kms; $\chi^2_{\rm{}red}=10$), suggesting that the \ion{N}{3} line
profile varies in a somewhat different manner than the \ion{Si}{4}
lines.  This conclusion remains even if one removes the five most
discrepant points, which occur near phase 0.5, when the \ion{N}{3}
velocity may be biased towards positive values by the blue-shifted
absorption feature of~H$\delta$.

The radial velocities inferred from the various regions in the red
spectra are much more discrepant.  Again, the velocities inferred from
regions dominated by lines from the same multiplet are more similar to
each other (for the two \ion{N}{2} regions: rms of 9~\kms;
$\chi^2_{\rm{}red}=5.4$), while the velocities inferred from the
\ion{Al}{3} line are more different (rms of 9.7~\kms;
$\chi^2_{\rm{}red}=11$). 

In summary, we find that for the blue regions, the velocities inferred
from the different lines track each other closely, but not exactly,
while for the red region the correlation is much less tight.  We
stress, though, that no systematic differences are present.  Indeed,
below we will find that for all regions consistent radial-velocity
amplitudes are inferred.  Also, the differences from the best-fit
radial-velocity curve will turn out to be substantially larger than
the differences between the different lines, reflecting the fact that,
to first order, the lines do behave similarly.

\section{The radial-velocity curve of HD~77581}\label{sec:velcurve}

Ideally, the radial velocities reflect only the motion of the
supergiant companion around the system's center of gravity, and are
described by a function characterizing a Keplerian orbit with known
period $P$, eccentricity $e$, periastron angle $\omega$,
radial-velocity amplitude $K_{\rm{}opt}$ and systemic velocity
$\gamma$.  We fitted the radial velocities derived for the different
lines, leaving as free parameters only $K_{\rm{}opt}$ and
$\gamma$\footnote{Since cross-correlation yields only relative
velocities, the fitted value of $\gamma$ does not reflect the true
systemic velocity of the binary.}, since the other parameters are
constrained very accurately by X-ray timing (Bildsten et al.\
\cite{BCC+97}; see Table~\ref{tab:batse}).

In practice, we are confronted with the fact that the individual
radial-velocity measurements deviate significantly from the best-fit
Keplerian orbit. This problem was already recognized in previous
studies of \VelaX1.  In Paper~II, we found that these velocity
``excursions'' are correlated on a time scale of a day and must be
non-orbital.  This is the reason why we chose for the observational
strategy to obtain one spectrum per night over many orbits of the
system, with the expectation that the velocity excursions would
average out, thus allowing an accurate determination of the
radial-velocity amplitude.

\begin{figure}
\centerline{\psfig{figure=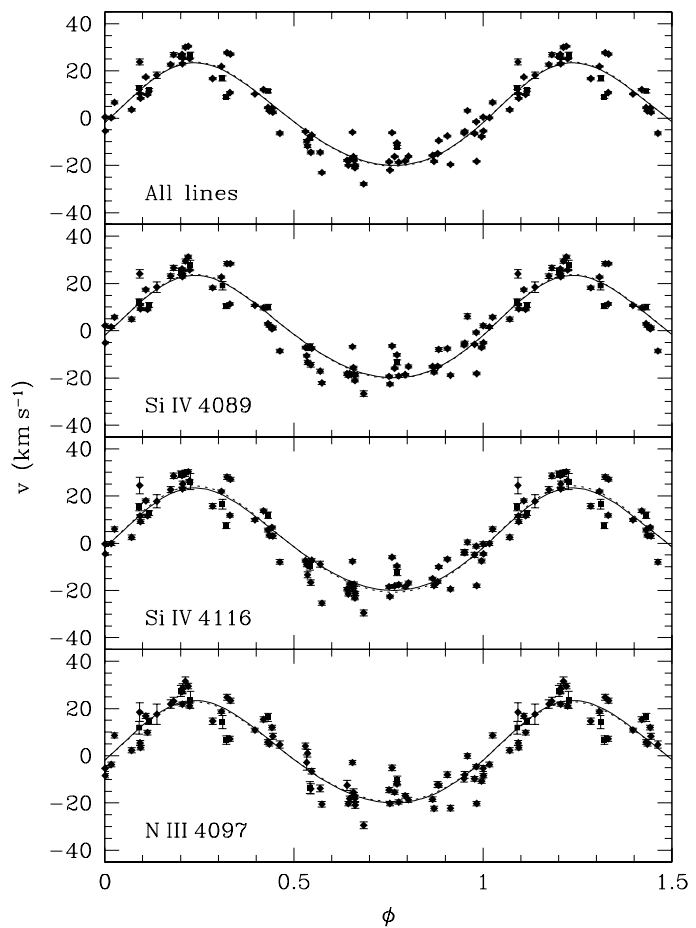,width=9cm}}
\caption[]{Radial-velocity curves derived from the two \ion{Si}{4} and
the \ion{N}{3} regions separately, and one derived from the three
regions combined (labelled ``All lines'').  The individual
radial-velocity measurements (filled circles) with their formal error
bars are shown together with the orbital fit to the curves (drawn
line; free parameters for the fit are the relative $\gamma$-velocity
and the radial-velocity amplitude $K$; see Table~\ref{tab:radvel}).
The dotted line reflects the best-fit curve of Fig.~\ref{fig:vcurveall}.}
\label{fig:vcurveblue}
\end{figure} 

\begin{figure}
\centerline{\psfig{figure=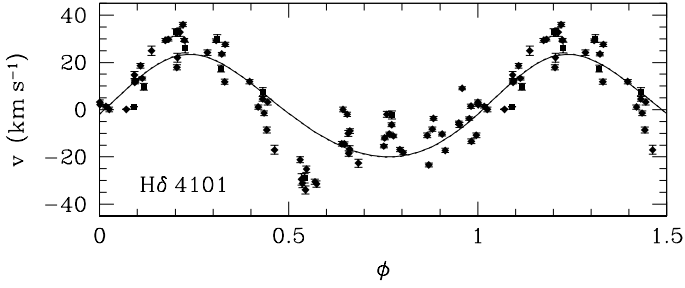,width=9cm}}
\caption[]{Radial-velocity curve derived from the H$\delta$ line.
Note the large deviation towards negative velocities around orbital
phase 0.5 due to the presence of a photo-ionisation wake.}
\label{fig:hdcurve}
\end{figure}

\begin{figure}
\centerline{\psfig{figure=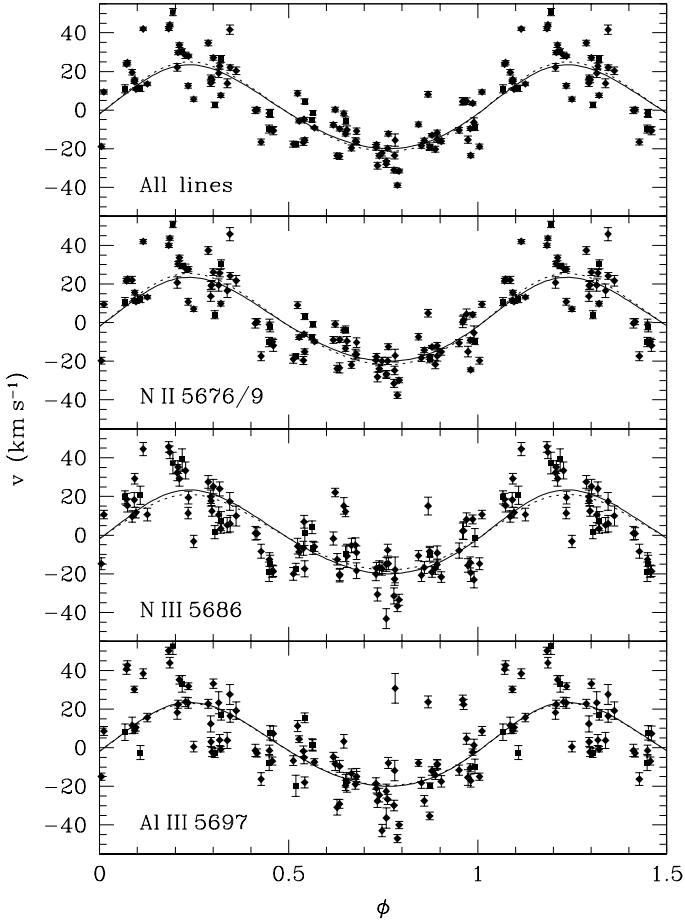,width=9cm}}
\caption[]{Radial-velocity curves derived from each region separately
and from all regions together for the red spectra.}
\label{fig:vcurvered}
\end{figure} 

In Fig.~\ref{fig:vcurveblue}, we show the radial velocities derived
from the two \ion{Si}{4} and the \ion{N}{3} regions separately, as
well as the radial-velocity curve derived from all these regions
combined.  Overlaid are the best-fit radial-velocity curves; the
fitted radial-velocity amplitudes are listed in
Table~\ref{tab:radvel}.  One sees that there are large random velocity
deviations from the Keplerian orbit, with root-mean-square deviations
of $\sim\!7~\kms$.  Superposed on these random velocity excursions,
there appear to be systematic deviations with orbital phase: the
velocities are mainly above the best-fit curve between phases 0.15 and
0.25 and between 0.7 and 0.9, and mainly below the fit between phases
0.45 and 0.65.  We will return to this in
Sect.~\ref{sec:systematics}.

The radial-velocity curve derived from the H$\delta$ line
(Fig.~\ref{fig:hdcurve}) shows much stronger systematic deviations
with orbital phase, in particular an on average large negative
velocity excursion between phases 0.45 and 0.65 (when the neutron star
passes in front of the supergiant).  As we discussed in
Sect.~\ref{sec:bluespec}, the H$\delta$ profile is strongly
distorted at these orbital phases, due to an additional blue
absorption component.  This component moves towards more negative
velocities, also distorting the \ion{N}{3} line profile.  This likely
is also the reason for the slight difference between the \ion{N}{3}
and \ion{Si}{4} radial-velocity curves.  We did not include the region
including H$\delta$ in the final determination of the amplitude
of the radial-velocity curve.

In Fig.~\ref{fig:vcurvered}, we show the radial-velocity curves
derived from cross-correlation of the \ion{N}{2} 5676, 5679~\AA, the
\ion{N}{2} 5686~\AA, and the \ion{Al}{3} regions, as well as of all
regions combined.  Fitted amplitudes are listed in
Table~\ref{tab:radvel}.  As mentioned, we did not use the \ion{C}{3}
5710.77 \AA\ region because of possible contamination with the diffuse
interstellar band.

From the figure, the scatter around the fitted radial-velocity curves
is much larger than in the blue, with root-mean-square deviations of
up to $20~\kms$ (Table~\ref{tab:radvel}).  On top of this large
scatter, systematic trends are harder to see, but in general the
behaviour seems consistent with that in the blue.

We will return to the effects of the possible systematic effects with
orbital phase below, after discussing the radial velocities inferred
from ultraviolet spectra, about which there has been some controversy,
and including these and other velocities in a combined velocity curve.

\section{Reanalysis of the IUE spectra}\label{sec:iue}

A large number of ultraviolet spectra of HD~77581 have been obtained
with the {\em International Ultraviolet Explorer} (IUE). In
Paper~II, we used these to determine radial velocities as well.  We
found relatively large uncertainties, of about 2.5~\kms, but given the
large amplitude of the intrinsic variability, they nevertheless were a
useful supplement to our optical observations.

After our study, a separate analysis of the IUE spectra was performed
by Stickland et al.\ (\cite{SLR97}).  These authors included a larger
set of spectra and determined velocities by cross-correlation with
spectra of $\kappa$~Ori rather than with spectra of HD~77581 itself.
They also used a somewhat different range of wavelength regions for
calculating the cross-correlation profiles.  The radial-velocity
amplitude they derived is significantly smaller than that found by us;
it is also inconsistent with that derived from optical spectra in
Papers~I \& II.  Clearly, in order to have confidence in our results,
it is necessary to understand the origin of this discrepancy.

We have contacted Drs.~Stickland and Lloyd and they kindly provided us
with the IUE spectra (reduced using their procedures), as well
as the procedures and wavelength regions used for normalization and
cross-correlation.  We found that the main cause for the discrepancy
was that in the Stickland et al.\ cross-correlation procedure for
determining the stellar velocity, regions around interstellar and wind
lines were zeroed in both programme-star and comparison-star spectra.
Combined with slight inadequancies in normalization, this led to the
introduction of artificial lines at fixed wavelengths in both spectra.
In consequence, the radial velocities are biased systematically
towards zero and the radial-velocity amplitude is underestimated.
Dr.~Stickland confirmed this conclusion, and reports that after
correction of the procedure, the radial-velocity amplitude increases
to 22~\kms\ (compared with the 17.8~\kms\ reported in their
publication), thus solving the discrepancy.

In order to provide what we hope will be definitive radial velocities
for the IUE spectra, we decided to re-do the complete analysis.
For this purpose, we retrieved all 52 SWP spectra of HD~77581 from the
archive.  These archive spectra have been processed using NEWSIPS, the
final IUE reduction pipeline.  We determined radial-velocity
differences between all pairs of images in the same way as done in
Paper~II, using the same wavelength regions (Table~4 in Paper~II), and
used these velocity differences to infer velocities for the individual
spectra.  Furthermore, we determined radial velocities relative to the
spectra of $\kappa$~Ori used by Stickland et al.\ (\cite{SLR97}):
SWP~7745 (small aperture) and SWP~13776 (large aperture).

We found that nine spectra of HD~77581 were of insufficient quality.
For eight of these, the background was abnormally high (SWP~19041,
19062, 25762, 25763, 25847, 25848, 25849, 25851), while for the
remaining one, the integration was very short (1~ks; SWP~25850).  We
have not used any of these further.

In our analysis, it became clear that many of the cross-correlation
profiles show small excess peaks.  These peaks, first found by Evans
(\cite{Ev88}) and discussed at some length in Paper~II (see Fig.~4 in
that paper) are thought to result from the presence of so-called
fixed-pattern noise in the detector.  Apparently, the new calibration
employed in NEWSIPS has not resolved this.  They occur only if the
spectra are taken through the same aperture.  As in Paper~II, we
corrected for such excess peaks by including an extra Gaussian peak in
the fit to the discrete cross-correlation profile (if spectra were
taken through the same aperture).

In Table~\ref{tab:iue} in the Appendix, we list the radial velocities
derived from the cross-correlations of all spectra with each other.
We verified that these are consistent with those derived from
cross-correlation with $\kappa$~Ori.  We prefer not to use the latter,
however, because $\kappa$~Ori has a smaller $v\sin{i}$ than HD~77581,
and thus is not a good template.  We note that Stickland et al.\
(\cite{SLR97}) argued that this actually makes it a better choice as a
template.  We do not agree: as one uses a template with narrower and
narrower lines, approaching delta functions, the cross-correlation
procedure reduces more and more to a weighted sampling of the line
profile.  Hence, the position of the peak of the cross-correlation
profile will become a measurement simply of the position of the
deepest part of the line profile.  In contrast, when one uses a
template that matches in line width, all information in the line
profile is used (in a way closely corresponding to $\chi^2$
minimalisation).  We stress, however, that our result does not depend
on this choice: from the radial velocities in Table~\ref{tab:iue}, we
find $K_{\rm{}opt}=21.0\pm1.6~\kms$ (Table~\ref{tab:radvel}), while
from the velocities determined by cross-correlation with the small and
large aperture spectra of $\kappa$~Ori, we derive
$K_{\rm{}opt}=22.5\pm2.0$ and $22.9\pm2.0~\kms$, respectively.
Similarly, our results would change very little if we chose the
wavelength regions used by Stickland et al.\ (\cite{SLR97}) instead of
the ones from Paper~II.

\begin{table}
\caption[]{Radial-velocity amplitudes for the different datasets.}
\label{tab:radvel}
\begin{tabular}{lllll}
\hline
\noalign{\smallskip}
Dataset & $N_{\rm d}/N_{\rm n}$\rlap{$^{\rm a}$}& rms\rlap{$^{\rm b}$}&  
Ampl.\rlap{$^{\rm c}$}& $K$\rlap{$^{\rm d}$}\\
\hline
\noalign{\smallskip}
Paper~I &                \pho82/\pho29&$\ldots$&$\ldots$&$21.8\pm1.2$\\[.3ex]
Paper~II, all data$^{\rm e}$&			        
                           \pho79/\pho50&\pho6.5&\pho9.0&$20.8\pm1.3$\\
\quad CCD                 &\pho40/\pho11&\pho4.5&   11.3&$21.9\pm3.1$\\
\quad Photographic        &\pho13/\pho13&\pho4.0&\pho6.8&$21.5\pm1.9$\\[.3ex]
Blue, all lines           &\pho77/\pho75&\pho7.0&\pho8.2&$22.0\pm1.0$\\
\quad\ion{Si}{4} 4089\AA  &             &\pho7.1&\pho8.3&$22.0\pm1.0$\\ 
\quad\ion{N}{3} 4097\AA   &             &\pho7.4&\pho8.8&$21.1\pm1.1$\\
\quad\ion{Si}{4} 4116\AA  &             &\pho7.2&\pho8.5&$22.5\pm1.0$\\[.3ex]
Red, all lines            &   104/\pho97&   10.1&   15.0&$23.1\pm1.5$\\
\quad\ion{N}{2} 5676, 5679\AA&          &   10.1&   14.8&$23.6\pm1.5$\\
\quad\ion{N}{2} 5686\AA   &             &   11.3&   18.2&$19.5\pm1.9$\\
\quad\ion{Al}{3} 5697\AA  &             &   11.8&   19.9&$21.2\pm2.0$\\[.3ex]
IUE                       &\pho43/\pho40&\pho9.0&\pho6.4&$21.0\pm1.6$\\[.3ex]
CCD, Phot., Blue, IUE     &  173/139   &\pho8.4&\pho8.3&$21.7\pm0.8$\\[.3ex]
\hline
\end{tabular}

\smallskip\noindent
$^{\rm a}$ Total number of spectra and number of nights.\\
$^{\rm b}$ Root-mean-square deviation (in \kms) from a Keplerian
curve, including both measurement uncertainties and intrinsic
deviations.\\
$^{\rm c}$ Standard deviation (in \kms) of the normally distributed
deviation amplitudes required to reproduce the rms and used to
estimate the uncertainty in the radial-velocity amplitude (see text).\\
$^{\rm d}$ Inferred radial velocity amplitude (in \kms). \\
$^{\rm e}$ Including the velocities derived in Paper~II for 26 IUE
spectra.  These are a subset of those analysed in \S\ref{sec:iue}.
\end{table}

\section{The combined radial-velocity curve and systematic effects
with orbital phase}\label{sec:allvel}

\begin{figure*}
\centerline{\psfig{figure=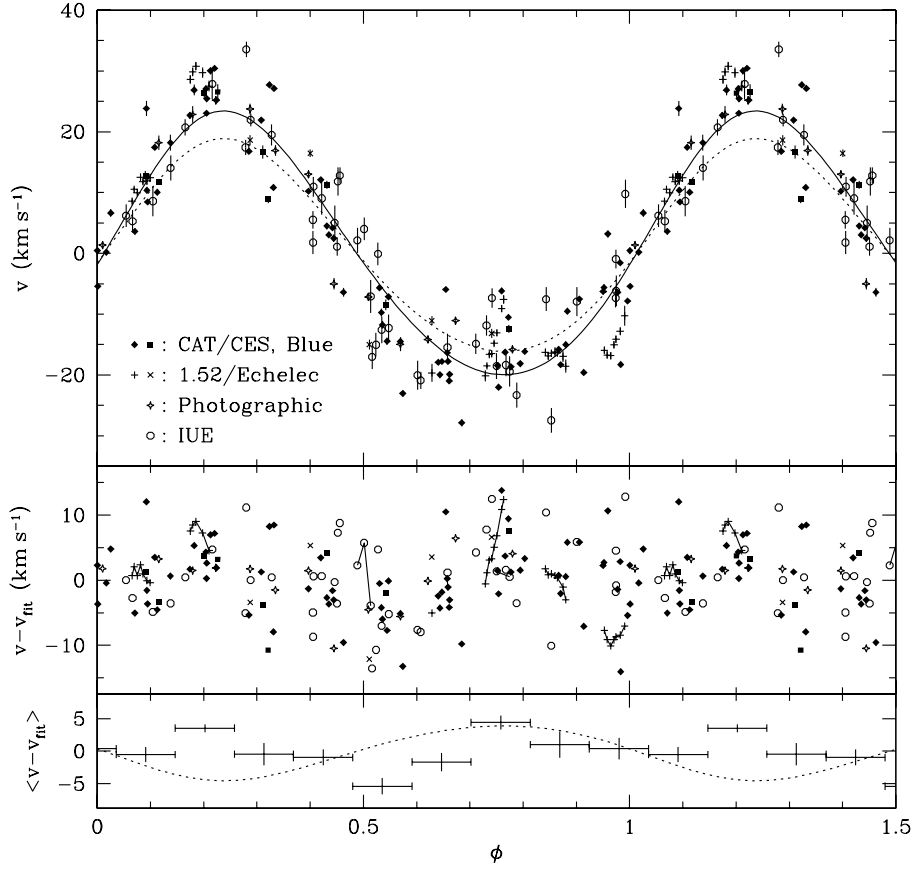,width=12cm}}\hfill
\caption[]{Radial-velocity curve derived from the combined data of
this paper and Paper~II (top panel). Overdrawn is the Keplerian curve
that best fits the nightly averages of the data (solid line), as well
as the curve expected if the neutron star has a mass of
$1.4$~$M_{\odot}$ (dotted line; $K_{\rm opt}=17.5~\kms$).  The
residuals to the best-fit are shown in the middle panel.  For clarity,
the error bars have been omitted.  Points taken within one night are
connected with lines.  In the bottom panel, the residuals averaged in
9 phase bins are shown.  The horizontal error bars indicate the size
of the phase bins, and the vertical ones the error in the mean.  The
dotted line indicates the residuals expected for a $1.4$~$M_{\odot}$
neutron star.}
\label{fig:vcurveall}
\end{figure*}

To obtain the best measure of the radial-velocity amplitude, we
combined the radial-velocity measurements of HD~77581 from our blue
dataset with those derived from the IUE analysis and those
obtained in Paper~II from 13 photographic spectra and 40 CCD echelle
spectra.  We do not use the velocities derived from our red spectra,
given their much larger scatter.  The resulting radial-velocity curve
is shown in Fig.~\ref{fig:vcurveall}.  The best-fit radial-velocity
amplitude is $K_{\rm opt}=21.7\pm0.8{\rm\,km\,s^{-1}}$.  In
Table~\ref{tab:radvel}, we compare this with the values inferred in
earlier sections from subsets of the data, as well as with those found
previously.  Clearly, the determinations are all consistent with each
other.

\subsection{Search for systematic effects with orbital phase}
\label{sec:systematics}

Our approach of obtaining radial velocities over an extended period of
time was motivated by the hope that for a given orbital phase the
large and highly significant radial-velocity deviations from a
Keplerian orbit would average out.  In other words, that just by
adding measurements with random deviations, the accuracy of the
derived radial-velocity amplitude, and thus the mass of the neutron
star, would be improved.  If there were to be {\it systematic}
deviations with orbital phase, however, the measured mean radial
velocity at a given orbital phase would not necessarily be equal to
the Keplerian radial velocity of the star in its orbit.

We can test the above assumption by verifying whether the residuals
binned over certain fractions of orbital phase are consistent with
zero.  In the lower panel of Fig.~\ref{fig:vcurveall}, we show the
mean of the residuals determined for 9 phase bins, as well as the
associated errors in the mean, calculated from the standard deviation
in the phase bins.  It is clear that these are not consistent with
zero.  The root-mean-square deviation of these phase-binned residuals
is $2.7~\kms$, while only $1.35~\kms$ is expected based on the
Monte-Carlo simulations.  From those simulations, the likelihood that
such a high value occurs by chance, under the assumption of no
orbital-phase related systematic effects, is smaller than 0.1\%.

Another piece of evidence that systematic, phase-locked effects occur,
is that the eccentricity one infers from the optical data is 0.18 (and
$\omega=355^\circ$).  From our simulations, the probability to find a
value this much larger than the BATSE value is less than~1\%.

Understanding the physical origin of these systematic deviations might
allow one to account for them and to further improve the accuracy of
the determination of the radial-velocity amplitude.  We discuss three
possible mechanisms below.

\subsection{Wind effects and evidence for the existence of a 
photoionisation wake}\label{sec:ionisationwake}

In the orbital phase range 0.45--0.65, strong velocity excursions are
observed in the radial-velocity curve obtained for the H$\delta$ line
(Sect.~\ref{sec:radvel}).  In this phase interval, the X-ray source
passes through the line of sight of the B supergiant. Inspection of
the H$\delta$ line shows that at these orbital phases the absorption
profile includes an additional blue-shifted absorption component,
similar to what is observed in time series of spectral lines formed in
the supergiant's stellar wind (e.g., Kaper et al. \cite{KHZ94}).  In
consequence, the cross-correlation procedure yields a more negative
radial velocity for the H$\delta$ line.

\begin{figure*}
\hbox to\textwidth{\hfill
\psfig{figure=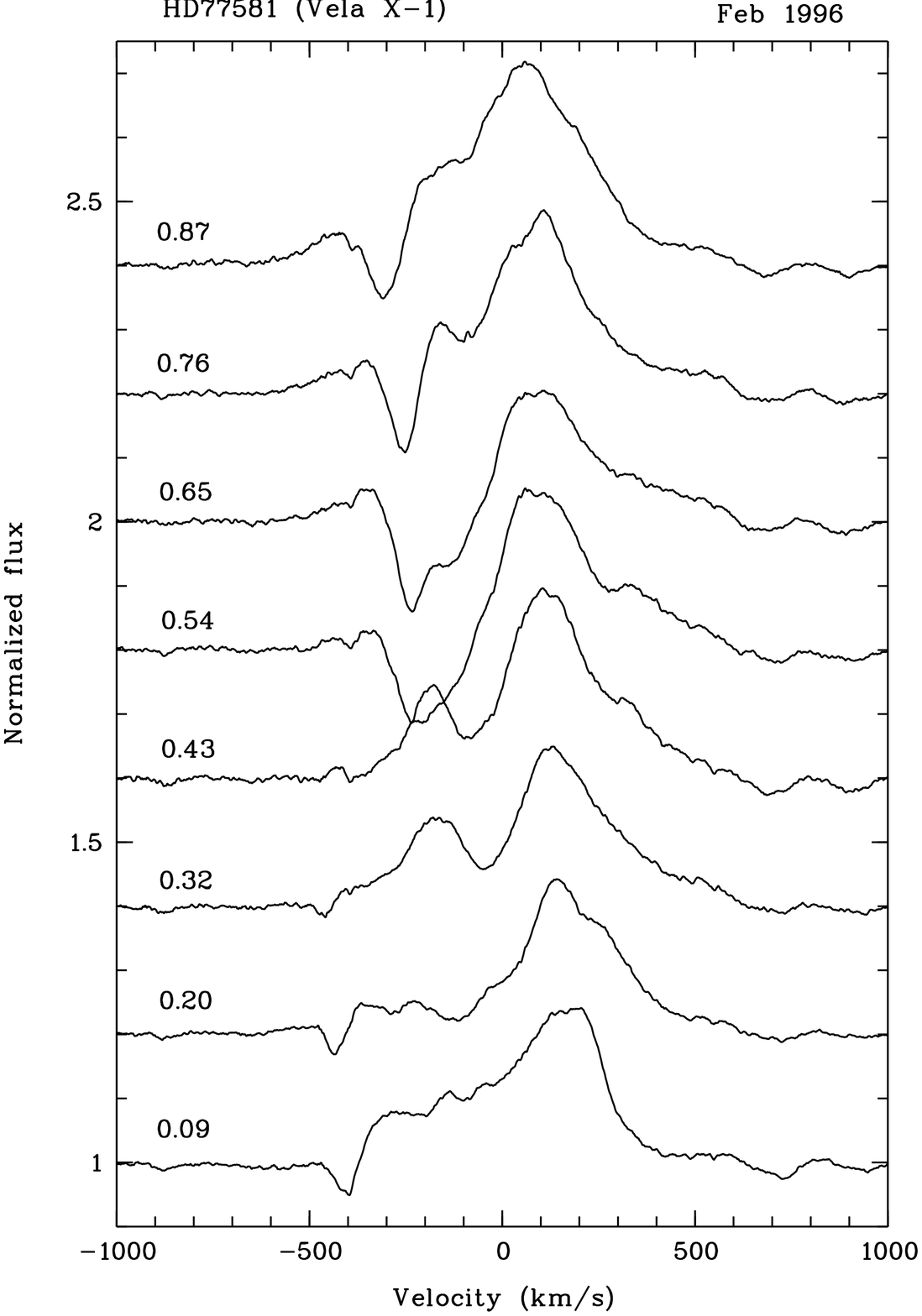,width=8cm,clip=}\hfill
\psfig{figure=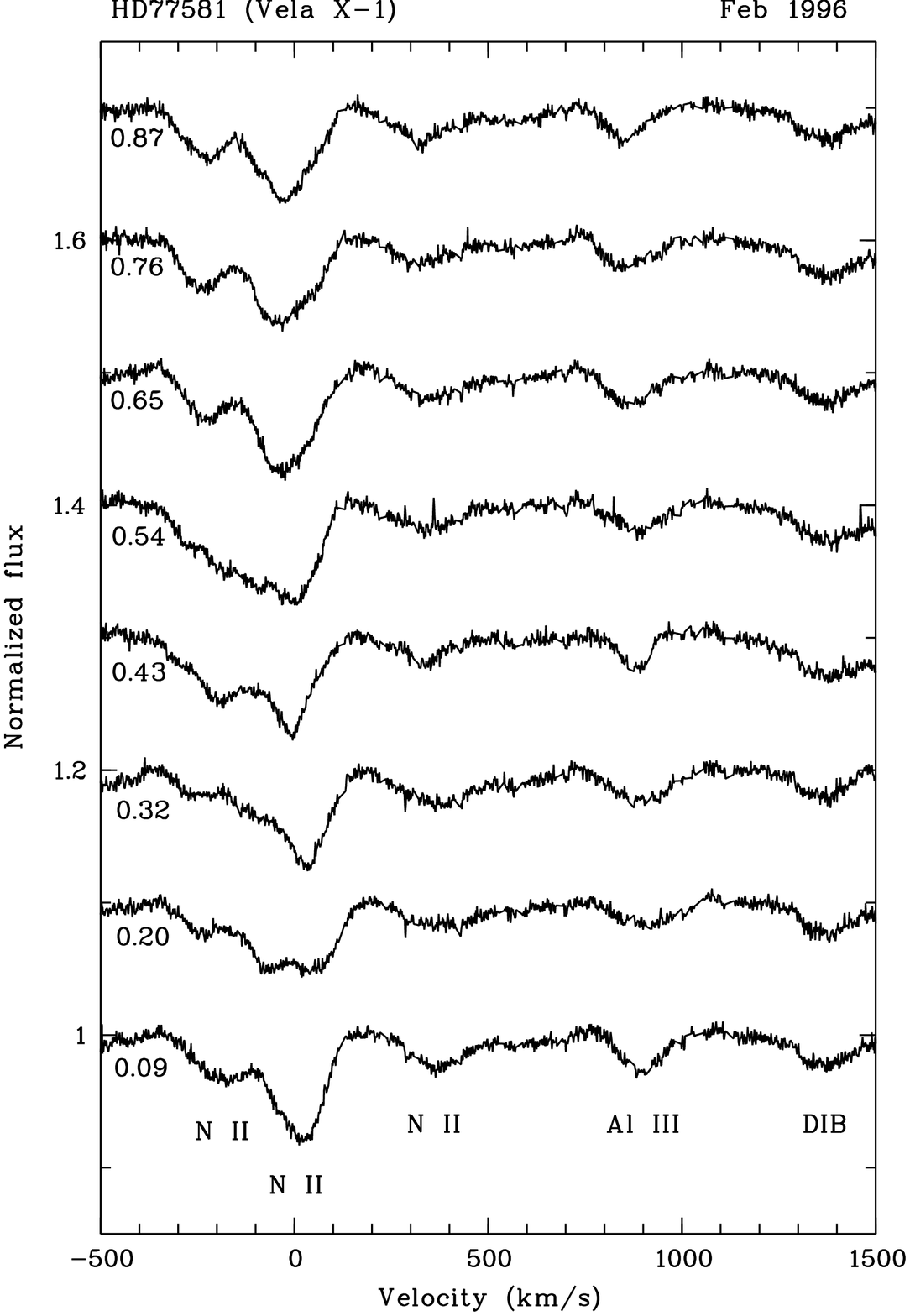,width=8cm,clip=}\hfill}
\caption[]{Sequence of H$\alpha$ profiles (left) obtained in February
1996, together with the simultaneously taken red spectra.}
\label{fig:feb96}
\end{figure*}

The origin of this additional blue-shifted absorption component is
likely a so-called photo-ionization wake.  The X-ray photons emitted
by the compact X-ray source fully ionize the surrounding wind regions,
creating an extended Str\"{o}mgren zone in the stellar wind.  The
presence of such a zone was predicted by Hatchett \& McCray
(\cite{HM77}), and can explain the observed strong orbital modulation
of the ultraviolet resonance lines of HD~77581 (e.g., Kaper et al.\
\cite{KH93}).  Due to the high level of ionization of the plasma
contained in the Str\"{o}mgren zone, inside this zone the radiative
acceleration of the stellar wind is quenched, leading to low wind
velocities within the Str\"{o}mgren zone.  Therefore, at the trailing
border of the Str\"{o}mgren zone a strong shock is formed where the
relatively fast ambient flow meets the slow wind that moved through
the Str\"{o}mgren zone.  The formation of such a ``photo-ionization
wake'' is clearly seen in hydrodynamical simulations (Blondin et al.\
\cite{BK90}).  Additional evidence for the existence of such a
structure was found in the X-ray light curve of \VelaX1 (Feldmeier et
al.\ \cite{FABN96}).

On two occassions, in February and May 1996, we obtained spectra of
strong optical lines formed in the stellar wind together with some of
the (red) spectra that have been used to measure the star's radial
velocity. In the period 10-18 February 1996 we monitored the H$\alpha$
line of HD~77581; Fig.~\ref{fig:feb96} shows the dramatic (wind)
variability observed in this line. At phase 0.54 a strong,
blue-shifted ($-200$ to $-300~\kms$) absorption component appears,
which at later phases migrates towards more negative velocities.
Another blue-shifted absorption component, most probably the remnant
of a previous passage of the X-ray source, is seen in the first
spectra obtained ($\phi=0.09$, 0.20, 0.32).  The emission component of
the H$\alpha$ profile is variable as well; notice the blue-shifted
emission component present at phase 0.3-0.4, most likely caused by
emission from the dense photo-ionization wake when the X-ray source is
approaching the line of sight. For comparison, Fig.~\ref{fig:feb96}
also displays the red spectra obtained within the same nights, where
zero velocity corresponds to the rest wavelength (in the heliocentric
frame) of the \ion{N}{2} line at 5679.56~\AA. This line profile
clearly demonstrates intrinsic variability which cannot be solely due
to orbital motion. However, apart from the enhanced blue-shifted
absorption at $\phi=$~0.54, there does not seem to be an obvious
correlation between the variability in the H$\alpha$ and the
\ion{N}{2} lines. Remarkable is the shallow (incipient emission?) and
broad (additional blue-shifted absorption component?) absorption
profile observed at phase~0.2.

\begin{figure*}
\hbox to\textwidth{\hfill
\psfig{figure=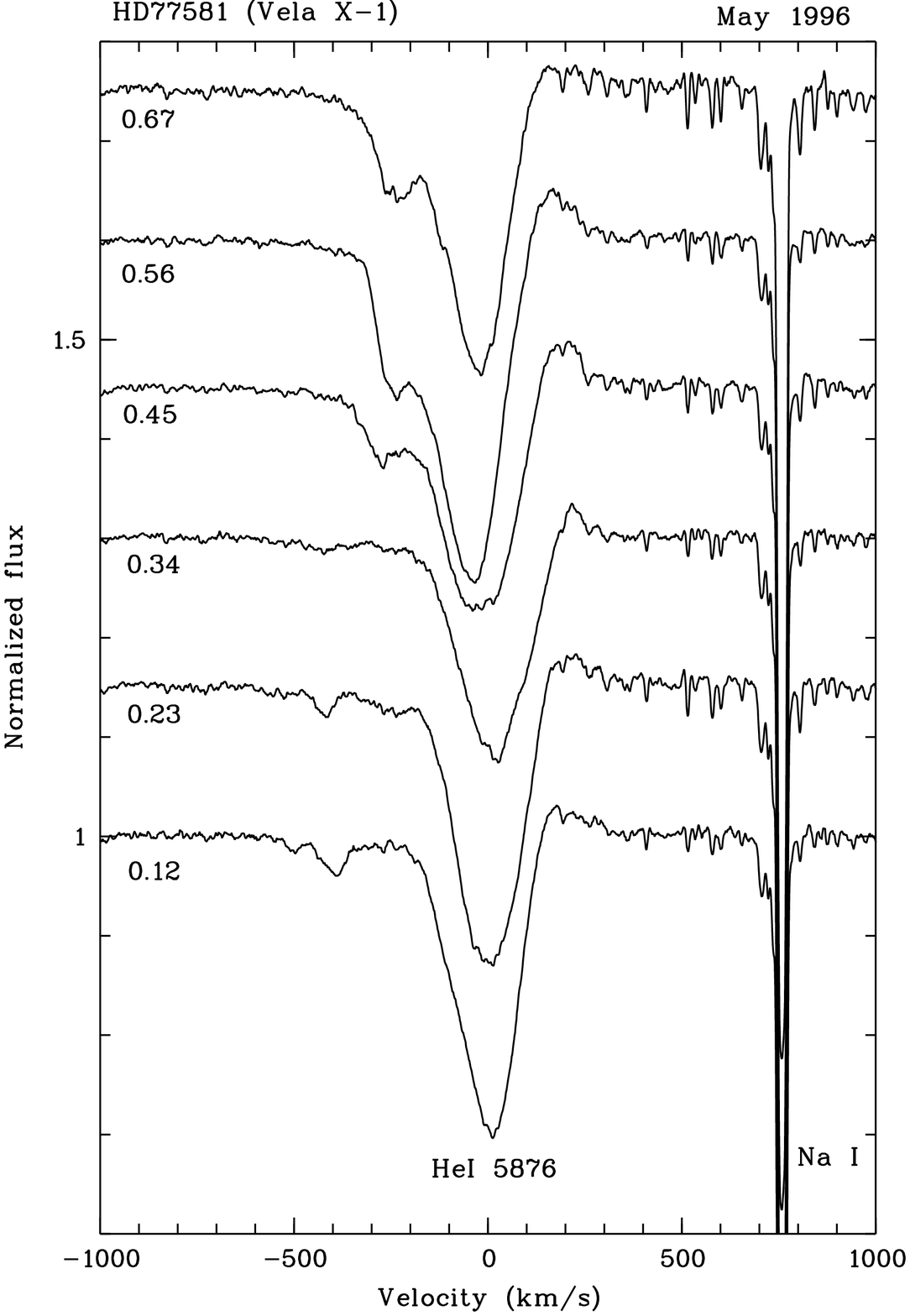,width=8cm,clip=}\hfill
\psfig{figure=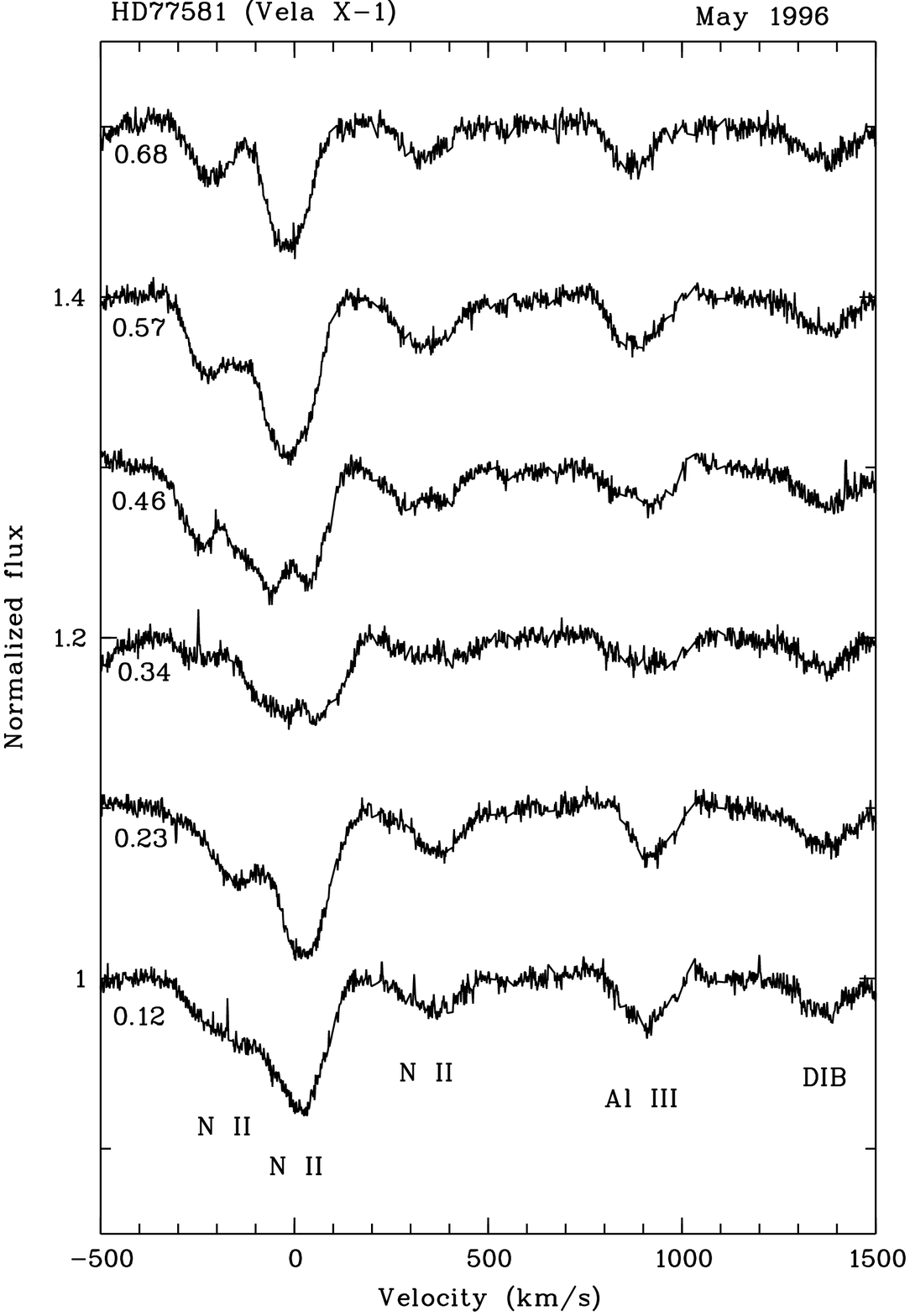,width=8cm,clip=}\hfill}
\caption[]{Sequence of He~{\sc i} profiles (left) obtained in May 1996
compared to the simultaneously taken red spectra.}
\label{fig:may96}
\end{figure*} 

From May 10 to 16, we monitored the \ion{He}{1} 5876~\AA\ line, which
is not as strong as the H$\alpha$ line, but is still sensitive to
changes in the base of the stellar wind. The systematic changes
(Fig.~\ref{fig:may96}) strongly resemble those observed in the
H$\beta$ line (Kaper et al. \cite{KHZ94}) and are similar to the
variations found in H$\alpha$. Comparison with the red \ion{N}{2}
spectra shows that at $\phi=0.34$ both the \ion{He}{1} and the
\ion{N}{2} (and \ion{Al}{3}) lines are shallower, while at $\phi=0.57$
all lines are deeper.  The blue-shifted absorption is enhanced at
phases 0.46 and~0.56.

We conclude that the systematic deviation in radial velocity of the
H$\delta$ line in the phase interval $0.45-0.65$ is caused by the
presence of a photo-ionization wake.  Weaker spectral lines are much
less strongly, though still measurably affected in this phase
interval; stronger lines like H$\alpha$ show large deviations in line
shape over a much wider range of the orbit. We decided to leave out
the radial-velocity measurements based on the H$\delta$ line in our
further analysis. In Sect.~\ref{sec:excludephases}, we investigate
the impact of the points between phase 0.45 and 0.65 on the
determination of the radial-velocity amplitude.  It turns out to be
small.

\subsection{Tidal distortion}
\label{sec:tidal}

Tidal deformation could also cause systematic deviations of the
radial-velocity curve, since for a deformed star `center-of-light'
radial-velocity measurements do not necessarily reflect the actual
center-of-mass velocity.  To investigate the effect this might have on
inferred orbital parameters, Van Paradijs et al.\ (\cite{vPTZ77}) made
model calculations for a star with properties appropriate for
HD~77581.  For simplicity, they assumed a circular orbit and
synchronous rotation of the primary, so that the shape of the star was
given by a single Roche equipotential surface at all phases.  Using
theoretically predicted line profiles, including the \ion{Si}{4} 4089
line, they find that appreciable systematic effects can occur,
differing from ion to ion, with the main expected deviations being a
positive one around orbital phase 0.1 and a negative one around
phase~0.9.  As a result, an apparent eccentricity may arise and the
radial-velocity amplitude may be overestimated, in unfavourable cases
by up to~30\%.

The systematic deviations we encounter in the radial-velocity curve
behave differently.  The larger deviations occur around orbital phases
0.2 and~0.8, and at both phases the deviations are directed towards
positive velocities.  Thus, it seems the effects of the tidal
deformation are not as large as predicted.  One should keep in mind,
however that the assumptions made by Van~Paradijs et al.\
(\cite{vPTZ77}) do not correspond to the actual situation in \VelaX1:
the orbit is eccentric and HD~77581 rotates sub-synchronously (by a
factor of about 2/3, Zuiderwijk \cite{Zu95}).  Given the
possibly large effects on the inferred parameters, it would be
worthwile to reconsider these issues.

\subsection{Non-radial pulsations}
\label{sec:nrp}

Apart from distorting the shape of the star, tidal forces can excite
(non-radial) pulsations.  With a neutron star in a close eccentric
orbit (the distance between the neutron star and the non-synchronously
rotating surface of HD~77581 is about half a stellar radius) one can
expect that tidal waves are excited at the surface of HD~77581 (e.g.,
Witte \& Savonije \cite{WS99}). The neutron star's orbital frequency
may resonate with certain non-radial modes, perhaps phase-locked with
the orbit.  The properties of these modes depend on the details of the
internal structure of the supergiant, which has undergone a complex
history of binary evolution, including a phase of mass transfer and a
nearby supernova explosion forming the neutron-star companion.

Radial and low-degree non-radial pulsations can change the shape of
photospheric absorption line profiles significantly (e.g., Vogt \&
Penrod \cite{VP83}).  This is due to local temperature variations
induced by the pulsations and/or local Doppler shifts related to the
3-dimensional pulsation velocity field.

The question is whether such tidally induced (or self-excited)
pulsations, when present in HD~77581, are detectable through
inspection of the detailed shape of the line profiles and, if so,
could result in systematic deviations in the radial-velocity curve.
According to Telting \& Schrijvers (\cite{TS97}), low-order pulsation modes
(with degree $\ell~<$~3) can lead to detectable radial-velocity
variations and changes in the shape of the line-profiles. For
instance, moment analyses of the spectral line profiles (Balona
\cite{Ba86}, Aerts et al. \cite{AP92}) have been successfully used to
detect low-order modes in single pulsators such as the $\beta$~Cephei
stars.

If the presence of non-radial pulsations is the physical cause for the
observed deviations in radial velocity, knowledge of the pulsation
mode(s) might provide a tool to ``correct'' the measured radial
velocities and to improve the accuracy of the determination of the
Keplerian orbit of HD~77581. However, a mode can only be identified if
the time sampling of the data is sufficient to resolve its periodicity
(in practice several spectra per hour and continuous -- 24 hours per
day -- coverage). The time sampling of the spectra used in this paper
(one spectrum per day) is poor; it is sufficient to cover the change
in radial velocity due to orbital motion, but likely insufficient to
identify pulsation modes.  Another disadvantage of the current dataset
is the limited spectral coverage; this excludes the possibility to
combine many different spectral lines in order to increase the
signal-to-noise ratio and to separate line changes induced by the
temperature- and velocity distribution associated with a given
pulsation.

A preliminary moment analysis performed on the line profiles comprised
in our spectra did not yield a useful result.  This is an indication
that any putatuve non-radial pulsation modes are not phase-locked to
the orbital motion and/or do not persist for a long period of time.  A
dataset obtained with better time sampling and spectral coverage is
required to further investigate the occurrence of non-radial
pulsations in the atmosphere of HD~77581.

\subsection{Orbital-phase intervals excluded}
\label{sec:excludephases}

\begin{figure}[ht!]
\centerline{\psfig{figure=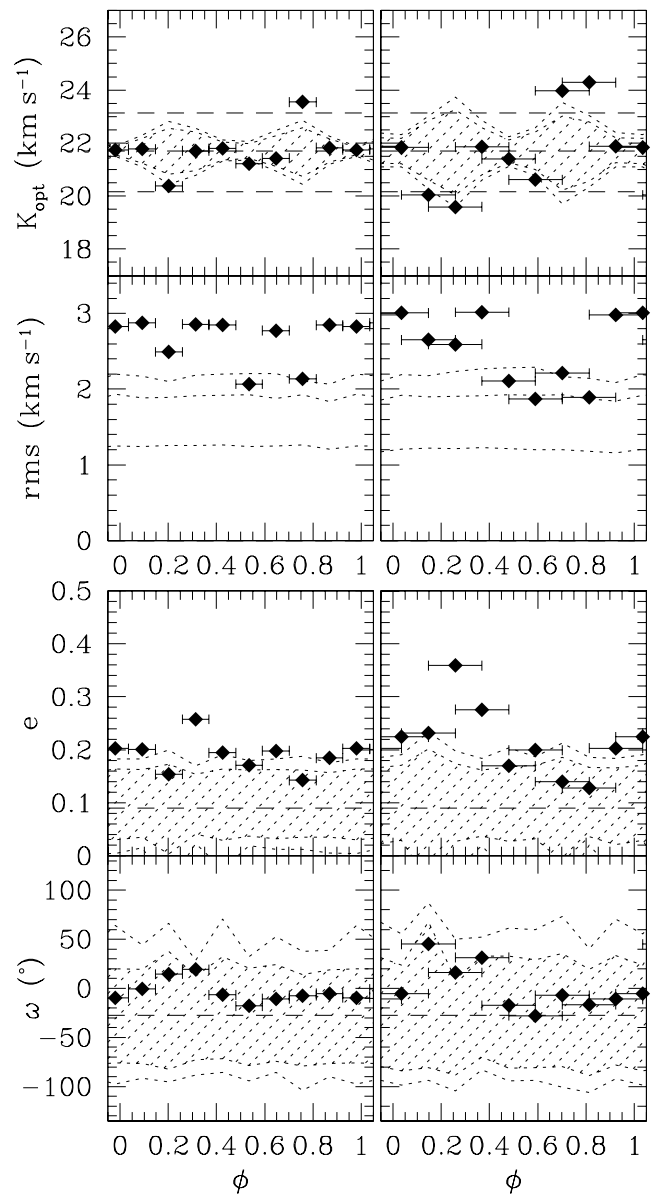,width=8.5cm}}
\caption[]{Results for fits to the radial-velocity curve with phase
intervals of width 1/9 (left) and 2/9 (right) excluded.  The points in
the top panel show the best-fit radial-velocity amplitudes found when
the phase range indicated by the horizontal error bars is excluded.
The short-dashed contours indicate the 95\% and 99\% confidence range
of the variation expected from our simulations (for which it is
assumed that the variability of the star is independent of orbital phase.)
The long-dashed lines indicate the best-fit amplitude to all data and
the associated 95\% confidence uncertainties.  The panels in the
second row show the root-mean-square deviation of the phase-binned
residuals.  The short-dashed lines indicate the levels below which
50\%, 95\%, and 99\% of the simulated data are contained.  In the
bottom two rows of panels, the best-fit values for the eccentricity
and periastron angle are shown.  The long-dashed lines indicate the
accurate values inferred from the BATSE measurements
(Table~\ref{tab:batse}), the short-dashed contours the 95\% and 99\%
ranges expected from the simulations.}
\label{fig:exclph}
\end{figure} 

In the H$\delta$ profile, we found clear evidence for an ionisation
wake in the phase interval between 0.45 and 0.65.  It may be that the
photoionisation wake has a (less obvious) effect on the other spectral
lines as well.  We can verify to what extent the exclusion of this
phase interval would affect the amplitude of the radial-velocity curve
(and thereby the measurement of the neutron-star mass).  Similarly, in
previous studies it was found that the largest deviations occurred
near velocity minimum (Paper~II and references therein).  From
Fig.~\ref{fig:vcurveall}, one sees that the phase bin at velocity
minimum again has one of the largest and most significant deviations
from zero.

In order to see what is the effect of these phase intervals on the
solution, we decided to exclude, like in Paper~II, all possible one
ninth and two ninth cycle wide orbital-phase intervals (i.e.,
approximately one and two-day wide) from the fit to the
radial-velocity curves, starting from orbital phase zero until one
with a step of one ninth.  Subsequently, we fitted the remaining
points and derived the amplitude of the radial-velocity curve, as well
as the uncertainty estimated using our Monte-Carlo simulations.  The
resulting amplitude is shown as a function of the excluded phase
intervals in Fig.~\ref{fig:exclph}.  Also shown is the rms of the
phase-binned residuals.  If a specific phase interval were responsible
for virtually all of the phase-locked deviation, one would expect that
the binned residuals would become consistent with zero when that phase
bin is excluded.  From the figure, one sees that excluding the bins
near inferior conjunction of the neutron star, when the ionisation
wake is most important, and near velocity minimum have the largest
effect on the rms.  Excluding inferior conjunction, the inferred
radial-velocity amplitude is not affected much, since the velocities
are at a zero crossing, while excluding velocity minimum, a much
larger velocity amplitude is inferred; indeed, excluding this bin
leads to the only change in radial-velocity amplitude whose
probability of occurring by chance, as inferred from the simulations,
is less than~1\%.

\subsection{Fitting eccentricity and periastron angle}
\label{sec:eomega}

In another attempt to determine whether any specific bin is
responsible for the phase-locked systematic effects, we repeated the
above simulations but with this time the eccentricity and the
periastron angle as free parameters in the fitting procedure.  In
principle, of course, one should reproduce the values found from the
X-ray analysis.  The bottom panels of Fig.~\ref{fig:exclph} show the
resulting eccentricity and periastron angle.  We find that when early
phase intervals are excluded, the inferred eccentricity remains far
larger than the true value, measured by Bildsten et al.\
(\cite{BCC+97}; Table~\ref{tab:batse}).  When phase intervals around
velocity minimum are excluded, however, the inferred eccentricity
becomes much closer to the true value.

\subsection{The  uncertainty in the radial-velocity amplitude}
\label{sec:bestestimate}

The above results suggest that the main systematic effects on the
radial-velocity curve occur near inferior conjunction and at velocity
minimum.  If so, the radial-velocity amplitude inferred from all the
data likely is an underestimate of the true value.  However, while we
believe the velocity deviation at inferior conjunction can be
understood -- in terms of a photo-ionisation wake -- we do not have a
good idea for the cause of the other systematic deviations.  Lacking
this, it is difficult to estimate the uncertainty that is introduced.
Probably best is to assume there is no bias to high or low
radial-velocity amplitude, and base the estimate on the increase in
the uncertainties required to match the observed root-mean-square
deviations of the phase-binned residuals.  This is about a factor two,
and corresponds to a similar increase in the uncertainty on the final
radial-velocity amplitude.  In consequence, our final estimate is not
much more precise than values given previously, in which the influence
of systematic effects with orbital phase was not taken into account.
We conclude that our best present estimate is
$K_{\rm{}opt}=21.7\pm1.6~\kms$.

\section{The mass of Vela~X-1}\label{sec:mass} 

From the ratio of the radial-velocity amplitudes of the optical and
X-ray components, we infer a mass ratio
\[
q \equiv M_{\rm X}/M_{\rm opt} = K_{\rm opt}/K_{\rm X} = 0.078\pm0.006,
\] 
after inserting the value of the radial-velocity amplitude of the X-ray
source (derived from X-ray pulse-timing analysis)
\[ K_{\rm X}\equiv (2\pi/P)a_{\rm{}X}\sin{}i(1-e^2)^{-1/2} =
278.1\pm0.3~\kms\ \, . \]  
Combined with the X-ray mass function
\[ 
f_{\rm X} \equiv \frac{M_{\rm X}^3\sin^3i}{(1+1/q)^2}
 = \frac{4\pi^2}{GP^2} (a_{\rm{}X}\sin{}i)^3 = 19.74\pm0.07\,M_\odot, 
\]
we infer 
\[ 
M_{\rm X}\sin^3i=1.78\pm0.15\,M_\odot, \, {\rm and}
\]
\[
M_{\rm opt}\sin^3i=22.9\pm0.3\,M_\odot. 
\]

The above numbers are lower limits to the actual masses, since the
inclination can only be smaller than $90^\circ$.  As in Paper~II, we
use the method introduced by Joss \& Rappaport (\cite{JR84}) to
estimate the inclination $i$, the semi-major axis $a$, the radius
$R_{\rm{}opt}$ and mass $M_{\rm{}opt}$ of the companion, the
co-rotation factor $f_{\rm co}$, and the neutron-star mass
$M_{\rm{}X}$, from the observed values of the orbital period $P$,
eccentricity $e$, periastron angle $\omega$, projected semi-major axis
of the neutron star $a_{\rm{}X}\sin{}i$, eclipse duration
$\theta_{\rm{}ecl}$, radial-velocity amplitude $K_{\rm{}opt}$, and
projected rotational velocity $v_{\rm{}rot}\sin{}i$ (see
Table~\ref{tab:batse}).  The corresponding uncertainties are found by
means of a Monte-Carlo error propagation technique, in which a large
number of trial evaluations are made for parameters drawn from the
observed uncertainty distribution (which are assumed to be normal,
except for $\theta_{\rm ecl}$, for which a uniform distribution is
used).  Following Joss \& Rappaport (\cite{JR84}), we assume that the
star is close to filling its Roche lobe, with a volume filling factor
between 90 and 100\% at periastron.

The results are listed in Table~\ref{tab:batse}; for the mass of the
neutron star, we find
\[
M_{\rm X}= 1.86\pm0.16\,M_\odot.
\]
Here we should add a note of caution: as in Paper~II, for a
substantial fraction (37\%) of the trials, the randomly-chosen
parameters are inconsistent with each other in the sense that the
eclipse duration cannot be reproduced for any inclination.  We discard
these trials, but note that as a result, the inferred parameters are
based on effective distributions of the input parameters which no
longer reflect the actual ones.  Most likely, this is because the
assumption that the star is in instantaneous hydrostatic equilibrium
is incorrect (see Paper~II for a more extensive discussion).

Given the above uncertainty, and given that for constraints on the
neutron-star equation of state, the most interesting result is a firm
lower limit to the mass of the neutron star, we prefer to give lower
limits in terms of $M_{\rm X}\sin^3i$.  For these, we use the
one-sided 95\% and 99\% confidence lower limits to the radial-velocity
amplitude, of $19.1$ and $17.9~\kms$, respectively, for which we infer
$M_{\rm{}X}\sin^3i>1.54$ and $1.43~M_\odot$, respectively.

\begin{table}
\caption[]{Parameters of the Vela~X-1/HD\,77581 system.}
\label{tab:batse}
\begin{tabular}{@{}llll@{}}
\hline
Parameter & Value$^{\rm a}$ & Unit & Ref.\\
\hline
\multicolumn{4}{l}{\em Observed ($1\sigma$ uncertainties)}\\
$ T_{\pi/2}{}^{\rm a}          $&$2448895.7186(12)$& JD&  Bildsten et
                                                          al.\ \cite{BCC+97}\\
$ P_{\rm orb}                  $&$ 8.964368(40)   $& day\\
$ a_{\rm X}\sin i              $&$ 113.89(13)     $& lt-s\\
$ e                            $&$ 0.0898(12)     $& \\
$ \omega_{\rm X}               $&$ 152.59(92)     $&$^\circ$\\
$ \theta_{\rm ecl}^{\rm b}     $&  30--36          &$^\circ$& See Paper I\\
$ v_{\rm rot}\sin i            $&$ 116(6)         $&\kms& Zuiderwijk 
                                                          \cite{Zu95}\\[2mm]
$ K_{\rm opt}                  $&$ 21.7(16)       $&\kms& This paper\\[2mm]
\multicolumn{4}{l}{\em Derived ($1\sigma$ uncertainties)}\\
$ M_{\rm X}\sin^3i             $&$ 1.78\pm0.15    $&$M_\odot$\\
$ M_{\rm opt}\sin^3i           $&$ 22.9\pm0.3     $&$M_\odot$\\[2mm]
\multicolumn{4}{l}{\em Inferred (95\% confidence ranges$^{\rm c}$)}\\
$ f_{\rm co} $&$ 0.69_{-0.08}^{+0.08}$&\\[1mm]
$ i          $&$ >\!73               $&$^\circ$\\[1mm]
$ a          $&$ 53.6_{-0.9}^{+1.7}  $&$R_\odot$\\[1mm]
$ R_{\rm opt}$&$ 30.4_{-2.1}^{+1.6}  $&$R_\odot$\\[1mm]
$ M_{\rm opt}$&$ 23.8_{-1.0}^{+2.4}  $&$M_\odot$\\[1mm]
$ M_{\rm X}  $&$ 1.86_{-0.32}^{+0.32}$&$M_\odot$\\[1mm]
\hline
\end{tabular}
\vskip1mm\noindent
$^{\rm a}$ Time of mean longitude $90^\circ$.\\
$^{\rm b}$ 99\% confidence range.\\
$^{\rm c}$ For the inferred quanties, 95\% confidence ranges are
listed, since the uncertainties on many parameters have highly
non-normal distributions.  The one exception is the mass of the
neutron star, whose uncertainty distribution is close to normal.
Hence, the associated $1\sigma$ error is about $0.16~M_\odot$.
\end{table}

\section{Ramifications}\label{sec:discussion}

The best estimate of the mass of \VelaX1 is $1.86~M_\odot$.
Unfortunately, no firm constraints on the equation of state are
possible, since systematic deviations in the radial-velocity curve do
not allow us to exclude a mass around $1.4~M_\odot$ as found for other
neutron stars.  Therefore, we will not discuss the equation of state
here, but only address evolutionary ramifications of what still most
likely is a massive neutron star, {\em assuming} that the equation of
state is sufficiently stiff for such a massive neutron star to exist.

One reason that the high mass is peculiar (if true), is that the most
accurately determined neutron-star masses fall in a surprisingly small
range: for the binary radio pulsars with neutron-star companions, the
masses are in the range 1.338--1.442~$M_{\odot}$.  Indeed, Thorsett \&
Chakrabarty (\cite{TC99}) found that all binary radio pulsars have
masses which are consistent with a normal distribution around
$1.35~M_\odot$, with the remarkably small spread of $0.04~M_\odot$.
Furthermore, for all X-ray binaries other than \VelaX1 for which
dynamical masses are available, the neutron-star masses are consistent
with the same value (Van Kerkwijk et al.\ \cite{vKvPZ95}).

Obviously, since for no equation of state masses substantially {\em
lower} than $1.4~M_\odot$ are excluded, at least the lower limit to
the above narrow range must be set by the way in which the neutron
star is formed, i.e., by the physics of supernova explosions and the
evolution of stars massive enough to reach core collapse.  This leads
one to wonder whether it could not be the formation mechanism on its
own that leads to the narrow range in mass, just as for white dwarfs
the formation leads to masses mostly within a very narrow range
around~$0.6~M_\odot$, well below the Chandrasekhar mass.

If the formation indeed leads to a narrow range of masses, high masses
could arise only if substantial amounts of matter are accreted.  This
is not expected to have occurred for any of the radio or X-ray pulsars
with accurate mass estimates.  For low-mass X-ray binaries, in which
substantial accretion is expected to have occurred, masses around
$2~M_\odot$ have been suggested (see, e.g., Orosz \& Kuulkers
[\cite{OK99}] for an analysis of Cyg~X-2, and Zhang et al.\
[\cite{ZSS97}] for inferences based on quasi-periodic oscillations).
All these estimates, however, rely to greater or lesser extent on
unproven assumptions.  Furthermore, for the putative descendants,
radio pulsars with white dwarf companions, there is no evidence for
such high masses (Thorsett \& Chakrabarty \cite{TC99} and references
therein).

From recent evolutionary calculations, Timmes et al.\ (\cite{TWW96})
indeed expect that massive stars in close binaries, which explode as
Type Ib supernovae, give rise to initial neutron star masses in a
narrow range around $1.3~M_\odot$.  This values does not include
subsequent mass accretion from a reverse shock or from a massive
component in a binary system, and Timmes et al.\ expect that the final
masses could be somewhat higher.  Interestingly, for single stars,
which explode as Type II supernovae, they found a bimodal distribution
of initial neutron-star masses, with narrow peaks at 1.27 and
$1.76~M_{\odot}$.  As mentioned, they did not find a bimodal
distribution for stars in close binaries, but at present it is not
clear whether this result will hold (Woosley 2000, private
communication).  If stars in close binaries turn out to be more
similar to single stars after all, one could assign most neutron stars
to the first peak, and \VelaX1 to the second.

We end by noting that even if neutron stars are formed with masses in
two peaks, it might be rather unlikely to find a massive neutron star
in a close double neutron star binary.  This is because such systems
require a common-envelope stage, in which a merger can only be avoided
if the initial orbit was very wide.  Stars massive enough to form a
massive neutron star, however, likely do not evolve through a
red-giant phase, and a common-envelope phase would occur only for
rather close orbits, for which the binary would merge.

\begin{acknowledgements}
The authors wish to thank the ESO La Silla Observatory staff for their
dedicated efforts in executing this difficult programme. Without the
help of the night assistants at the CAT (and sometimes of the
observers themselves whose programme got interrupted) this programme
would not have been possible. A special word of thanks goes to Jesus
Rodriguez, the night operator at ESO Headquarters who carried out the
remote part of the observations from Garching. We thank the referee,
Dr.\ Phil Charles, for carefully reading the manuscript and for his
suggestions that helped to improve the paper. OB acknowledges the
support of an ESO studentship. LK and MHvK are supported by
fellowships of the Royal Netherlands Academy of Arts and Sciences.
\end{acknowledgements}

\appendix

\section{Log of observations}

The tables in this Appendix listing the log of observations of the
blue CES spectra (Tab.~\ref{tab:blue}), the red CES spectra
(Tab.~\ref{tab:red}), and the IUE spectra (Tab.~\ref{tab:iue}) are
only available in electronic form at the CDS via anonymous ftp to
cdsarc.u-strasbg.fr (130.79.128.5) or via
http://cdsweb.u-strasbg.fr/cgi-bin/qcat?J/A+A/. To provide an
impression of the information contained in these tables, only the
first five lines of each table are displayed here. 

\begin{table*}[p]
\caption[]{Observing log and radial velocities for the blue spectra
(central wavelength 4102~\AA).  The second digit of the identification
(first column) specifies the CCD used (b4 for CCD\#34 and b8 for
CCD\#38).  In the second column the barycentric-corrected Julian Date
of each observation is listed, and in the third the orbital phase
calculated using the ephemeris of Bildsten et al.\ (\cite{BCC+97}; see
Table~\ref{tab:batse}).  The following four columns list radial
velocity and associated uncertainty (in \kms) for each spectral region
separately, as derived by means of cross-correlation.  The last
columns lists radial velocity and uncertainty derived from the
cross-correlation of the ensemble of the two \ion{Si}{4} and the
\ion{N}{3} regions.}
\label{tab:blue}
{\small
\begin{tabular}{cccr@{$\pm$}lr@{$\pm$}lr@{$\pm$}lr@{$\pm$}lr@{$\pm$}l}
\hline
Ident & BJD$_{\rm mid}$&$\phi_{\rm orb}$& 
\multicolumn{2}{c}{\ion{Si}{4}~4089}&
\multicolumn{2}{c}{\ion{N}{3}~4097}&
\multicolumn{2}{c}{H$\delta$~4102}&
\multicolumn{2}{c}{\ion{Si}{4}~4116}&
\multicolumn{2}{c}{Si/N ensemble}\\ 
\hline
b8001& 2450006.8514&  0.9502&$ -5.99$&  0.82&$ -9.60$& 1.47&$ -5.71$& 0.77&$ -3.87$& 1.05&$ -6.27$& 0.51\\
b8002& 2450006.8660&  0.9518&$ -5.31$&  0.83&$ -8.05$& 1.43&$ -6.52$& 1.06&$ -3.89$& 1.04&$ -5.63$& 0.51\\
b8004& 2450008.8590&  0.1741&$ 23.11$&  0.97&$ 21.90$& 1.75&$ 29.18$& 0.81&$ 22.66$& 1.26&$ 22.69$& 0.58\\
b8005& 2450009.8532&  0.2850&$ 18.14$&  0.71&$ 14.68$& 1.36&$ 24.07$& 1.15&$ 15.72$& 1.04&$ 16.79$& 0.45\\
b8006& 2450010.8543&  0.3967&$ 10.76$&  0.43&$ 10.87$& 0.81&$ 11.75$& 0.94&$  9.90$& 0.62&$ 10.23$& 0.31\\
\hline
\end{tabular}}
\end{table*}

\begin{table*}[p]
\caption[]{Observing log and radial velocities of the red spectra
(central wavelength 5965~\AA).  Columns as in Table~\ref{tab:blue}.}
\label{tab:red}
{\small
\begin{tabular}{lllr@{$\pm$}lr@{$\pm$}lr@{$\pm$}lr@{$\pm$}lr@{$\pm$}l}
\hline
Ident & BJD$_{\rm mid}$&$\phi_{\rm orb}$&
\multicolumn{2}{c}{\ion{N}{2} 5676/9}&
\multicolumn{2}{c}{\ion{N}{2} 5697}&
\multicolumn{2}{c}{\ion{Al}{3} 5686}&
\multicolumn{2}{c}{\ion{C}{3} 5705}&
\multicolumn{2}{c}{N/Al ensemble}\\ 
\hline 
r8002r&  2450006.8545& 0.9505&$-10.55$& 1.83&$ -7.98$& 3.97&$-11.61$& 2.62&$ -2.47$& 4.25&$-10.48$& 1.38\\
r8001r&  2450012.8401& 0.6182&$ -8.96$& 1.82&$ -1.77$& 3.46&$ -4.75$& 2.45&$ -1.56$& 3.78&$ -7.65$& 1.29\\
r8003r&  2450013.8483& 0.7307&$-19.48$& 1.88&$-20.12$& 3.48&$-18.55$& 2.93&$-11.60$& 4.74&$-19.25$& 1.38\\
r8004r&  2450013.8629& 0.7323&$-17.66$& 1.72&$-17.13$& 2.95&$-21.05$& 2.39&$ -9.17$& 3.32&$-17.92$& 1.22\\
r8005r&  2450014.8585& 0.8433&$ -7.36$& 1.59&$-10.61$& 2.69&$ -7.93$& 1.79&$ -1.39$& 3.77&$ -7.43$& 1.16\\
\hline
\end{tabular}}
\end{table*}

\begin{table*}[p]
\caption[]{IUE observations of HD~77581.}
\label{tab:iue}
\begin{tabular}{llllr@{}}
\hline
SWP&    JD$_{\rm mid.~exp.}$& Exp.~time& Orbital& 
   \multicolumn{1}{l}{Velocity$^{\rm b,c}$}\\
number& $-$2440000&           (min.)&    phase$^{\rm a}$&
\multicolumn{1}{l}{(~km~s$^{-1}$)}\\
\hline
\pho 1442& 3628.708& 180& 0.451&$   0.0\pm1.5$\\
\pho 1488& 3634.118& 150& 0.055&$   5.1\pm1.9$\\
\pho 2087& 3712.994& 125& 0.854&$ -28.6\pm2.0$\\
\pho 2390& 3745.203& 108& 0.447&$   3.9\pm2.8$\\
\pho 3499& 3843.592& 120& 0.422&$   7.9\pm2.7$\\[1mm]
\hline
\end{tabular}
\vskip1mm\noindent
$^{\rm a}$ Using the ephemeris of Bildsten et al.\ \cite{BCC+97}
(see Table~\ref{tab:batse}).\\
$^{\rm b}$ Relative to SWP\,1442; corrected for shift of interstellar lines.\\
$^{\rm c}$ Quoted are 1$\sigma$ errors.\\
\end{table*}

\end{document}